\documentclass[cernpreprint, atlasdraft=false, texlive=2023, UKenglish, texmf, orcidlogo]{atlasdoc}
\PreprintIdNumber{CERN-EP-2022-207}
 
\usepackage{atlaspackage}
\usepackage{atlasbiblatex}
 
\usepackage{atlasphysics}
 
\usepackage{siunitx}
\ExplSyntaxOn
\cs_new_eq:NN \calc \fp_eval:n
\ExplSyntaxOff

\usepackage{setspace}
 
\usepackage{amsmath}
\usepackage{tikz-feynman}
\tikzset{
photon/.style={decorate, decoration={snake}, },
electron/.style={},
antielectron/.style={},
gluon/.style={decorate,
decoration={coil,amplitude=2.5pt, segment length=3pt}},
higgs/.style={dashed,}
}
 
\graphicspath{{logos/}{figures/}}
 
\usepackage{ANA-HDBS-2019-24-PAPER-defs}
\usepackage{mathrsfs}

\AtlasTitle{Search for a light charged Higgs boson in $t\to H^{\pm}b$ decays, with $H^{\pm} \to cb$, in the lepton+jets final state in proton-proton collisions at $\sqrt{s}=13$ TeV with the ATLAS detector}
 
\AtlasVersion{2.2}
\AtlasAbstract{
A search for a charged Higgs boson, $H^{\pm}$, produced in top-quark decays, $t \to H^{\pm}b$, is presented.
The search targets $H^{\pm}$ decays into a bottom and a charm quark, $H^{\pm} \to cb$.
The analysis focuses on a selection enriched in top-quark pair production, where one top quark decays into a leptonically decaying $W$ boson and a bottom quark,
and the other top quark decays into a charged Higgs boson and a bottom quark.
This topology leads to a lepton-plus-jets final state, characterised by an isolated electron or muon
and at least four jets. The search exploits the high multiplicity of jets containing $b$-hadrons,
and deploys a
neural network classifier that uses the kinematic differences between the signal and the background.
The search uses a dataset of proton-proton collisions collected at a centre-of-mass energy $\sqrt{s}=13$ TeV between 2015 and 2018 with the ATLAS detector at CERN's Large Hadron Collider, amounting to an integrated luminosity of 139~fb$^{-1}$.
Observed (expected) 95\% confidence-level upper limits between
0.15\% (0.09\%) and 0.42\% (0.25\%)
are derived for the product of branching fractions $\mathscr{B}(t\rightarrow H^{\pm}b) \times \mathscr{B}(H^{\pm}\rightarrow cb)$
for charged Higgs boson masses between 60 and 160~\gev, assuming the SM production of the top-quark pairs.}
 
\author{The ATLAS Collaboration}
 
\AtlasRefCode{HDBS-2019-24}

\PreprintIdNumber{CERN-EP-2022-207}

\arXivId{2302.11739}
 
\HepDataRecord{https://www.hepdata.net/record/ins2635801}

\AtlasJournalRef{JHEP 09 (2023) 004}
\AtlasDOI{10.1007/JHEP09(2023)004}

\hypersetup{pdftitle={ATLAS document},pdfauthor={The ATLAS Collaboration}}
 
\begin{document}
 
\maketitle
 
\tableofcontents
 
\FloatBarrier
\section{Introduction}

The Higgs boson's discovery by the ATLAS and CMS Collaborations at the Large Hadron Collider (LHC)~\cite{HIGG-2012-27,CMS-HIG-12-028} and subsequent campaigns to
precisely measure its properties~\cite{ATLAS:2022vkf,CMS:2022dwd} confirmed that the Standard Model (SM) of particle physics is an effective description of nature up to the \TeV\ energy scale.
 
Within the SM framework, the Brout--Englert--Higgs mechanism~\cite{Englert:1964et,Higgs:1964ia,Higgs:1964pj,Guralnik:1964eu,Higgs:1966ev,Kibble:1967sv}
is responsible for generating the masses of the gauge bosons via
electroweak symmetry breaking (EWSB). The Higgs boson emerges from the EWSB as the only physical spin-0 CP-even particle of the SM,
while the remaining components of the complex Higgs field doublet are absorbed into the longitudinal components of the gauge bosons.

In scenarios beyond the Standard Model, the Higgs sector is typically extended to incorporate new degrees of freedom.
A popular and minimal extension of the SM paradigm is provided by
two-Higgs-doublet models (2HDM)~\cite{Branco:2011iw}, where the Higgs sector consists of two
complex doublets, a mixture of the two doublets fulfils the same role as the SM Higgs field and generates a Higgs boson ($h$) similar to that in the SM, and the other mixture gives rise to a neutral CP-even Higgs boson ($H$), a neutral CP-odd Higgs boson ($A$), and a charged Higgs boson (\hp).
Unlike the SM, a general 2HDM allows flavour changing neutral current (FCNC) interactions at tree level, which need to be suppressed.
This is normally achieved by requiring that all fermions with the same electric charge couple to one Higgs doublet only~\cite{Glashow:1976nt,Paige:1977nz}, a condition referred to as ``Natural Flavour Conservation" (NFC).
Depending on the assignment of up/down-type quark and lepton couplings to each Higgs doublet, 2HDMs are categorised into four different types: type-I, type-II, type-III (lepton-specific), and type-IV (flipped, also known as type-Y).
A concrete realisation of a type-II 2HDM with NFC includes the Minimal Supersymmetric Standard Model (MSSM)~\cite{Fayet:1976et,Fayet:1977yc}. Similar Higgs sectors also arise in axion models~\cite{Kim:1986ax}.
 
Direct searches for new scalar particles, constraints from flavour observables, and precision measurements of the discovered Higgs boson suggest
that the natural mass scale for additional Higgs bosons from 2HDMs with NFC lies above several hundred \gev, depending on the assumed model parameters (see e.g. Refs.~\cite{Akeroyd:2016ymd,Kling:2020hmi,HIGG-2018-57,CMS-HIG-17-031}).
On the other hand, such constraints can be evaded or mitigated in other non-minimal extensions of the SM Higgs sector.
A particularly rich phenomenology is expected in models with three Higgs doublets (3HDM)~\cite{Akeroyd:2016ssd}, which
feature three CP-even and two CP-odd neutral Higgs bosons, as well as two charged Higgs bosons.
In 3HDMs the lightest charged Higgs boson can be lighter than the top quark
and can decay mainly into either a $\tau$-lepton and a neutrino, a charm quark and a bottom quark, or a strange quark and a charm quark~\cite{Akeroyd:2018axd}.
 
A search for $H^{\pm}\rightarrow cb$ decays\footnote{Charge conjugation is implied throughout the paper; the notation $cb$ is used in place of $c\bar{b}$/$\bar{c}b$.} in top-quark decays was performed by the CMS Collaboration~\cite{CMS-HIG-16-030}
using $19.7$~fb$^{-1}$ of proton--proton ($pp$) collision data collected at \sqrteight;
it reported upper limits at 95\% confidence level (CL) on the branching fraction $\BR(t\rightarrow H^{\pm}b)$ of (0.8--0.5)\%,
assuming $\BR(H^{\pm}\rightarrow cb) = 1.0$, for a charged Higgs boson mass (\hpm) between 90 and 150~\gev.
Related searches for \hpdeccs in top-quark decays were performed by the ATLAS~\cite{HIGG-2012-10} and CMS~\cite{CMS-HIG-13-035}
Collaborations, based respectively on 4.7~fb$^{-1}$ and 19.7~fb$^{-1}$ of $pp$ collision data collected at the centre-of-mass energies \sqrtseven and \sqrteight;
these searches reported upper limits at 95\% CL on the branching fraction $\BR(t\rightarrow H^{\pm}b)$ varying between 1\% and 5\%,
assuming $\BR(H^{\pm}\rightarrow cs) = 1.0$, for \hpm between 90 and 160~\gev.
 
\begin{figure}[htbp]
\begin{center}
\includegraphics[width=0.42\textwidth]{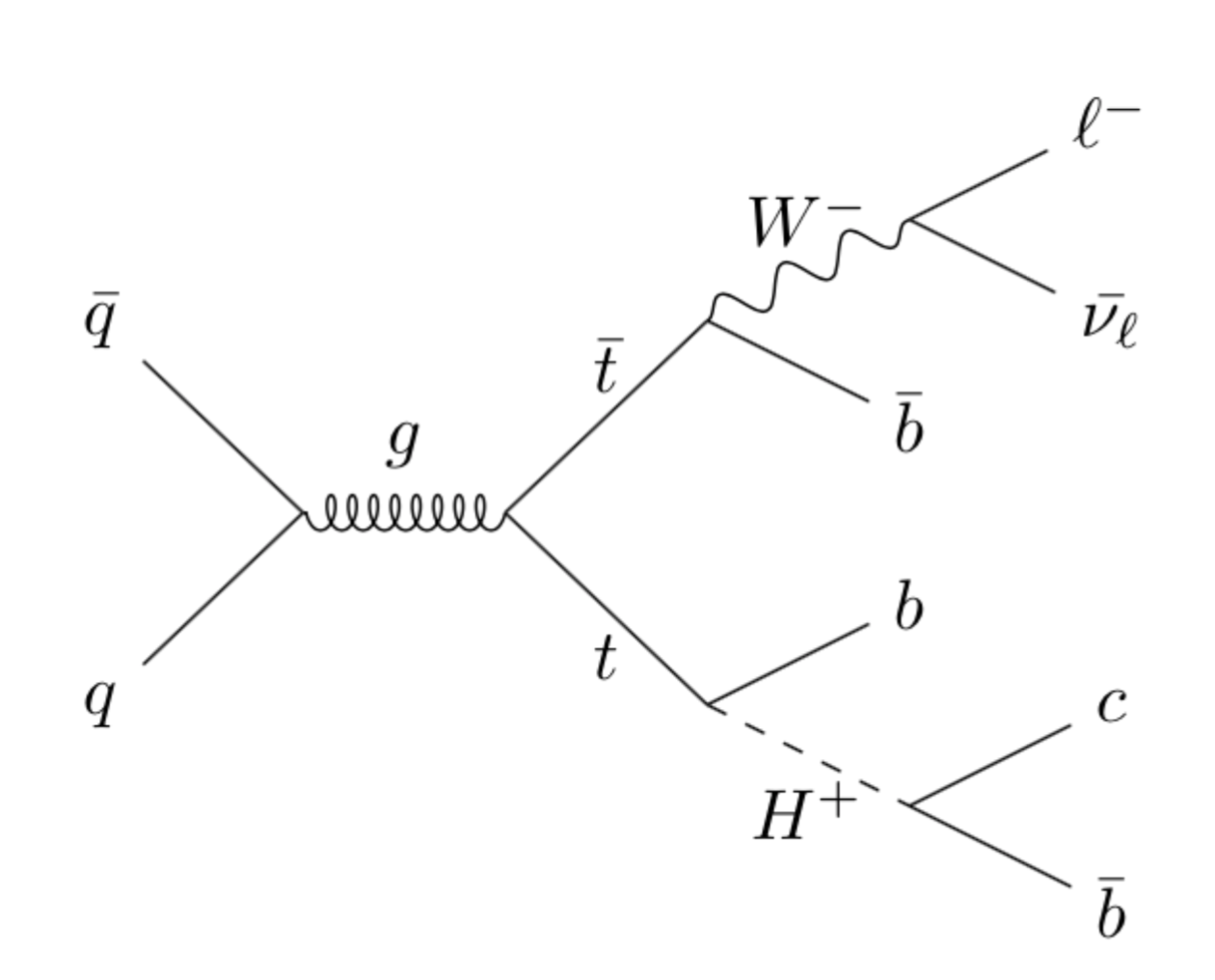}
\includegraphics[width=0.42\textwidth]{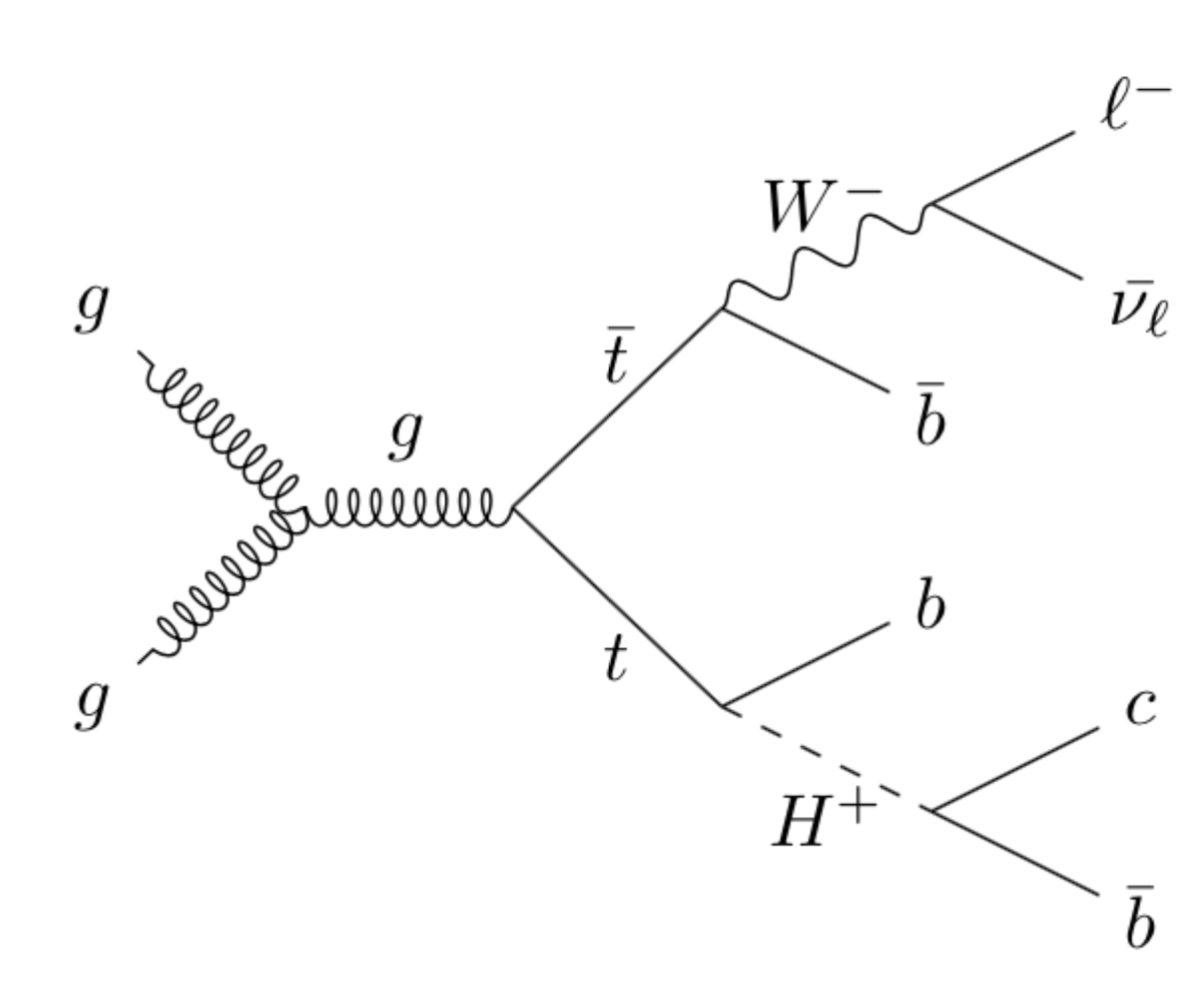} \\
\caption{Illustrative leading-order Feynman diagrams for the signal considered. Charge-conjugated Feynman diagrams are also assumed.}
\label{fig:diagram}
\end{center}
\end{figure}

This paper presents a search for $H^{\pm}\rightarrow cb$ in top-quark decays which is based on a dataset of $pp$ collisions collected at a centre-of-mass energy $\sqrt{s}=13\;\tev$ between 2015 and 2018 with the ATLAS detector, amounting to an integrated luminosity of 139~fb$^{-1}$.
The analysis focuses on a data sample enriched in top-quark pair production ($\ttbar$), where one top quark decays into a leptonically decaying $W$ boson and a bottom quark, and the other top quark decays into a $H^{\pm}$ boson and a bottom quark, as illustrated in Figure~\ref{fig:diagram}.
Compared to searches for \hpdeccharm\ and \hpdectau in \ttbar\ events,
searches for \hpdec\ take advantage of a significantly smaller yield of the irreducible SM background originating from \ttbar\ production with a $W$ boson decaying into
the Cabibbo--Kobayashi--Maskawa suppressed mode $W\rightarrow cb$.
The search exploits the high multiplicity of jets containing $b$-hadrons ($b$-jets), as expected from signal events, and deploys a neural network classifier that uses the kinematic differences between the signal and the background.
Thanks to a sizeably larger dataset and improved analysis techniques,
this search improves the expected sensitivity to \hpdec\ in top-quark decays by a factor of five compared to the previous publication~\cite{CMS-HIG-16-030} and explores an extended \hpm\ range, between 60 and 160~\gev.
 
\FloatBarrier
\section{ATLAS detector}

\newcommand{\AtlasCoordFootnote}{
ATLAS uses a right-handed coordinate system with its origin at the nominal interaction point (IP)
in the centre of the detector and the \(z\)-axis along the beam pipe.
The \(x\)-axis points from the IP to the centre of the LHC ring,
and the \(y\)-axis points upwards.
Cylindrical coordinates \((r,\phi)\) are used in the transverse plane,
\(\phi\) being the azimuthal angle around the \(z\)-axis.
The pseudorapidity is defined in terms of the polar angle \(\theta\) as \(\eta = -\ln \tan(\theta/2)\).
Angular distance is measured in units of \(\Delta R \equiv \sqrt{(\Delta\eta)^{2} + (\Delta\phi)^{2}}\).}

The ATLAS experiment~\cite{PERF-2007-01} at the LHC is a multipurpose particle detector
with a forward--backward symmetric cylindrical geometry and a near \(4\pi\) coverage in
solid angle.\footnote{\AtlasCoordFootnote}
It consists of an inner tracking detector surrounded by a thin superconducting solenoid
providing a \SI{2}{\tesla} axial magnetic field, electromagnetic and hadron calorimeters, and a muon spectrometer.
The inner tracking detector covers the pseudorapidity range \(|\eta| < 2.5\).
It consists of silicon pixel, silicon microstrip, and transition radiation tracking detectors.
Lead/liquid-argon (LAr) sampling calorimeters provide electromagnetic (EM) energy measurements
with high granularity.
A steel/scintillator-tile hadron calorimeter covers the central pseudorapidity range (\(|\eta| < 1.7\)).
The endcap and forward regions are instrumented with LAr calorimeters
for both the EM and hadronic energy measurements up to \(|\eta| = 4.9\).
The muon spectrometer surrounds the calorimeters and is based on
three large superconducting air-core toroidal magnets with eight coils each.
The field integral of the toroids ranges between \num{2.0} and \SI{6.0}{\tesla\metre}
across most of the detector.
The muon spectrometer includes a system of precision tracking chambers and fast detectors for triggering.
A two-level trigger system is used to select events.
The first-level trigger is implemented in hardware and uses a subset of the detector information
to accept events at a rate below \SI{100}{\kHz}.
This is followed by a software-based trigger that
reduces the accepted event rate to \SI{1}{\kHz} on average
depending on the data-taking conditions.
An extensive software suite~\cite{ATL-SOFT-PUB-2021-001} is used for real and simulated data reconstruction
and analysis, for operation and in the trigger and data acquisition systems of the experiment.


\FloatBarrier
\section{Objects definition and event selection}

Data from  $pp$ collisions at \sqrtonethree were recorded by the ATLAS detector between 2015 and 2018.
Only data for which all detector subsystems were operational are used~\cite{ATLAS:2019fst}; this dataset corresponds to an integrated luminosity of
139~\ifb~\cite{ATLAS-CONF-2019-021,Avoni:2018iuv}.
 
Events were recorded with a single-electron or single-muon trigger, with minimum thresholds for the transverse momentum (\pt) varying from 20 to 26~\gev\
depending on the lepton flavour and the data-taking period.
The triggers with the lowest \pt thresholds included isolation requirements based on the inner tracking detector or
electromagnetic calorimeter measurements~\cite{Aad:2020uyd,Aad:2019wsl,Aad:2020wji,ATLAS:2021lws}.
 
In each event, the primary vertex is defined as the reconstructed vertex having the highest scalar sum of the squared \pt of
associated tracks with $\pt \geq 0.5$~\gev.
 
Electrons are reconstructed from energy clusters in the electromagnetic calorimeter that are geometrically matched to a track in the inner tracking detector.
Electrons are required to satisfy $|\eta| < 2.47$ and a \enquote{Tight} identification requirement using  a likelihood-based method~\cite{ATLAS:2019qmc};
electrons are rejected if their calorimeter clusters lie within the transition region between the barrel and endcaps of the electromagnetic calorimeter,
$1.37 < |\eta| < 1.52$.
Muons are reconstructed from muon spectrometer tracks matched to tracks in the inner tracking detector
in the pseudorapidity range $|\eta| < 2.5$. Muon identification is based on \enquote{Medium} requirements~\cite{ATLAS:2020auj}.
Selected electrons and muons are required to have $\pt \geq 27$~\gev.
 
Isolation criteria are applied to the selected electrons and muons.
For electrons, the scalar sum of the transverse energy in calorimeter energy clusters within a cone of size
$\Delta R = 0.2$ around the electron is required to be less than 6\% of the electron \pt, excluding clusters originating from the electron itself.
For muons, the scalar sum of the \pt of tracks within a fixed-size cone around the muon (excluding its associated track)
must be less than 6\% of the muon \pt. The track isolation cone size is $\Delta R = 0.3$ for muon $\pt<50$~\gev\ and $\Delta R = 0.2$ for muon $\pt>50$~\gev.
 
Jets are reconstructed from topological energy clusters in the calorimeter~\cite{Aad:2016upy} using the particle-flow method~\cite{Aaboud:2017aca},
based on the \antikt clustering algorithm~\cite{Fastjet,Cacciari:2008gp} with a radius parameter of $0.4$.
The jet energy is calibrated at particle level~\cite{ATLAS:2020cli}, and jets are required to have $|\eta|< 2.5$ and a minimum \pt of 25~\gev.
For jets with $|\eta| <  2.4$ and $\pt < 60$~\gev, those originating from additional $pp$ collisions in the same or neighbouring bunch crossings (\pileup)
are suppressed by the use of the \enquote{jet-vertex-tagger}~\cite{ATLAS:2015ull}.
 
Jets containing $b$-hadrons are identified with the DL1r $b$-tagging algorithm~\cite{ATLAS:2022qxm}.
A jet is $b$-tagged if the DL1r score is above a certain threshold, referred to as an operating point (OP). Four OPs are defined with average
expected efficiencies for $b$-jets of 60\%, 70\%, 77\% and 85\%, as determined in simulated \ttbar\ events. The DL1r $b$-tagging score is divided into five exclusive bins according to the OPs.
The distribution obtained by ordering these five bins from higher to lower $b$-jet efficiency is referred to as the \enquote{pseudo-continuous} $b$-tagging score.
 
The missing transverse momentum, with magnitude \met, is defined as the negative vector sum of the transverse momenta of all selected and calibrated physics
objects. Low-momentum tracks from the primary vertex that cannot be associated with any of the reconstructed physics objects
described before are also included in the \met calculation~\cite{ATLAS:2018txj}.
 
A sequential overlap removal procedure is applied to ensure that the same calorimeter energy deposit or the same track is not associated with two or more different reconstructed objects, following the prescription described in Ref.~\cite{ATLAS:2020bhu}.
 
The events are required to have exactly one selected electron or muon that matches the lepton that fired the trigger, and at least four jets.  At least two of the jets are required to pass the 60\% OP $b$-tagging requirement and an additional jet is required to pass the 70\% $b$-tagging OP. In order to suppress background from multijet production, additional requirements are made on $\met$ as well as on the transverse mass of the lepton and $\met$ system ($\mtw$):\footnote{$\mtw = \sqrt{2 p^\ell_{\mathrm T} \met (1-\cos\Delta\phi)}$, where $p^\ell_{\mathrm T}$  is the transverse momentum (energy) of the muon (electron) and $\Delta\phi$ is the azimuthal angle separation between the lepton and the direction of the missing transverse momentum.} $\met >20~\gev$ and $\met +\mtw>60~\gev$. The above requirements are referred to as the ``preselection''.


\FloatBarrier
\section{Monte Carlo samples}
\label{sec:mc}

Monte Carlo (MC) simulation samples are used to model all backgrounds as well as the \hp\ signal, and evaluate related modelling uncertainties.
The main background for this search originates from  \ttbar production in association with jets,
followed by smaller contributions from single-top-quark, $V$+jets, \ttV, \ttH, diboson and other rare processes involving the production of a top quark.
Background due to non-prompt leptons is expected to be negligible, based on studies of data using multiple lepton isolation criteria~\cite{ATLAS:2021upq}
and analysis of low-\met\ events.
 
The matrix-element calculations for all generated samples use the \NNPDF[3.0nlo]~\cite{Ball:2012cx} set of parton distribution functions (PDFs), unless stated otherwise.
In all samples interfaced to \PYTHIA[8]~\cite{Sjostrand:2007gs} or \HERWIG[7]~\cite{Bahr:2008pv,Bellm:2015jjp},
the decays of bottom and charm hadrons were simulated using the \EVTGEN[1.2.0] program~\cite{Lange:2001uf}.
\PYTHIA[8.230]~\cite{Sjostrand:2014zea} modelled the parton shower,
hadronisation, and underlying event, with parameters set according
to the A14 tune~\cite{ATL-PHYS-PUB-2014-021} and using the \NNPDF[2.3lo]~\cite{Ball:2012cx}
PDF set; \HERWIG[7.04] used the H7UE tune and the
\MMHT[lo]~\cite{Harland-Lang:2014zoa} PDF set; and \HERWIG[7.13] used the
\HERWIG[7.1] default set of tuned parameters and the same PDF set.
The effect of \pileup was modelled by overlaying each simulated hard-scatter event with
inelastic $pp$ events generated with \PYTHIA[8.186] using the \NNPDF[2.3lo]~\cite{Ball:2014uwa} PDF set and the A3 tune~\cite{ATL-PHYS-PUB-2016-017}.
 
The generated events were processed through either a simulation~\cite{Aad:2010ah} of the ATLAS detector geometry and response
using \textsc{Geant4}~\cite{Agostinelli:2002hh} or a faster simulation, where the full \textsc{Geant4} simulation of
the calorimeter response is replaced by a detailed parameterisation of the shower shapes~\cite{FastCaloSim}.
Simulated events were processed through the same reconstruction software as the data, and corrections were applied so that the object identification efficiencies, energy scales
and energy resolutions matched those determined from data control samples.
 
\subsection{\ttbar background simulation}
 
The production of $\ttbar \rightarrow WbWb$ events (denoted simply $\ttbar$ in the following) was modelled using the
\POWHEGBOX[v2]~\cite{Frixione:2007nw,Nason:2004rx,Frixione:2007vw,Alioli:2010xd}
generator at next-to-leading order (NLO) with the \hdamp parameter\footnote{The
\hdamp parameter is a resummation damping factor and one of the
parameters that controls the matching of \POWHEG matrix elements to
the parton shower and thus effectively regulates the
high-\pt radiation against which the \ttbar system recoils.} set
to 1.5 times the top-quark mass, \mtop~\cite{ATL-PHYS-PUB-2016-020}.
The matrix-element calculation includes diagrams with a $b$-quark in the initial state using the five-flavour scheme~\cite{Maltoni:2012pa}.
The events were interfaced to \PYTHIA[8.230]~\cite{Sjostrand:2014zea} to model the parton shower, hadronisation, and underlying event.
 
The impact of using a different parton shower and hadronisation model was evaluated
by comparing the nominal \ttbar sample with another sample produced with the
\POWHEGBOX[v2] generator interfaced to \HERWIG[7.04] instead of \PYTHIA[8.230].
 
To assess the uncertainty associated with the NLO generator,
the \POWHEGBOX[v2] sample was compared with a sample of events
generated with \MGNLO[2.6.0] interfaced to
\PYTHIA[8.230].
 
Analogously to similar searches performed previously in ATLAS~\cite{ATLAS:2018jqi,ATLAS:2021upq}, the simulated $\ttbar$ events are categorised according to the flavour content of additional jets not originating from the decay of the $\ttbar$ system.
Events that have at least one $b$-jet, excluding heavy-flavour jets from top-quark or $W$-boson decays,
are labelled as \ttbarb; those with no additional $b$-jets but at least one charm-jet ($c$-jet) are labelled as \ttbarc; finally, events not containing
any additional heavy-flavour jets are labelled as \ttbarlight.
 
An additional sample to evaluate systematic uncertainties in the modelling of the \ttbarb\ process
was produced with the \POWHEGBOXRES~\cite{Jezo:2018yaf} generator and \OPENLOOPS~\cite{Buccioni:2019sur,Cascioli:2011va,Denner:2016kdg}, using a pre-release
of the implementation of this process in \POWHEGBOXRES provided by its authors~\cite{ttbbPowheg}.
It was interfaced to \PYTHIA[8.240].
The four-flavour scheme was used with the $b$-quark mass set to 4.95\,\GeV.
The factorisation scale was set to $0.5\times\Sigma_{i=t,\bar{t},b,\bar{b},j} m_{\mathrm{T},i}$, where $m_{\mathrm{T},i}=(m_{i}^{2}+p_{\text{T},i}^{2})^{1/2}$ and $j$ denotes extra light quarks or gluons,
the renormalisation scale was set to $\prod_{i=t,\bar{t},b,\bar{b}} m_{\mathrm{T},i}^{1/4}$,
and the \hdamp parameter was set to $0.5\times\Sigma_{i=t,\bar{t},b,\bar{b}}m_{\mathrm{T},i}$.
 
All generated \ttbar\ samples assume a diagonal Cabibbo--Kobayashi--Maskawa matrix, thus neglecting \ttbar\ events with rare $W\rightarrow cb$ decays.
Since such decays mimic the expected topology of the \hpdec\ signal, dedicated \ttbar samples were produced, setting $V_{cb}=0.041$.
A nominal sample was generated by using \POWHEGBOX[v2] at NLO with
the \hdamp parameter set to $1.5\times m_{\text{top}}$, and using \MADSPIN~\cite{Frixione:2007zp,Artoisenet:2012st}
to generate $W\rightarrow cb$ decays. The events were interfaced
to \PYTHIA[8.230] to model the parton shower, hadronisation, and underlying event.
Additional samples to assess the uncertainty in the
parton shower and hadronisation model and the NLO generator
were produced with \POWHEGBOX[v2] interfaced to \HERWIG[7.13] and  \MGNLO[2.6.0] interfaced to
\PYTHIA[8.230].
 
The \ttbar sample was normalised to the cross-section prediction at next-to-next-to-leading order (NNLO)
in QCD including the resummation of next-to-next-to-leading logarithmic (NNLL) soft-gluon terms calculated using
\TOPpp[2.0]~\cite{Beneke:2011mq,Cacciari:2011hy,Baernreuther:2012ws,Czakon:2012zr,Czakon:2012pz,Czakon:2013goa,Czakon:2011xx}.
For $pp$ collisions at a centre-of-mass energy of \rts~=~\SI{13}{\TeV}, this cross-section corresponds to
$\sigma(\ttbar) = \mathrm{832\pm51~pb}$ using a top-quark mass of $\mtop = 172.5\,\GeV$.
 
\subsection{Signal simulation}
 
Samples of $t\bar{t}\rightarrow H^{\pm}bW^{\mp}b$ events were generated using the \POWHEGBOX[v2] generator at NLO with
the \hdamp parameter set to $1.5\times \mtop$, and using \MADSPIN\ and \PYTHIA[8.230] to perform the top quark and \hpdec\ decays and to model the parton shower, hadronisation, and underlying event.
The $W$ boson was forced to decay leptonically to all three lepton flavours.
A total of 11 signal MC samples were generated with \hpm\ ranging from 60 to 160~\gev\ with 10~\gev\ spacing; the \hp\ boson's total width is assumed to be 1~\gev, more than ten times smaller than the expected mass resolution.
The signal samples were normalised to the same cross-section as used for the \ttbar background sample, and assuming an arbitrary product of branching fractions $\BR_{\text{ref}} = \BR(t\rightarrow H^{\pm}b)\,\times \BR(H^{\pm}\rightarrow cb) =1\%$.
 
\subsection{Other samples}
 
The associated production of top quarks with $W$ bosons ($tW$) and single-top-quark production in the $t$-channel and
$s$-channel were modelled by the \POWHEGBOX[v2]
generator at NLO in QCD using the five-flavour scheme.
The $tW$ process was modelled using the diagram removal scheme~\cite{Frixione:2008yi,Re:2010bp} to
handle interference and overlap with \ttbar production.
A related uncertainty was estimated by comparison with an alternative sample
generated using the diagram subtraction scheme~\cite{Frixione:2008yi,Re:2010bp}.
The events were interfaced to \PYTHIA[8.230].
The uncertainty due to the parton shower and hadronisation model was
evaluated by comparing the nominal samples of events with samples where
events generated with \POWHEGBOX[v2]  were interfaced to \HERWIG[7.04] instead of \PYTHIA[8.230].
To assess the uncertainty associated with the NLO generator,
the nominal samples were compared with samples generated
with \MGNLO[2.6.2] at NLO in QCD using the five-flavour
scheme and the \NNPDF[2.3nlo] PDF set. These events were
interfaced to \PYTHIA[8.230].
 
For the $tW$ channel single-top-quark process, the inclusive cross-section was normalised to the theory prediction
calculated at NLO in QCD with NNLL soft-gluon
corrections~\cite{Kidonakis:2010ux,Kidonakis:2013zqa}. The inclusive cross-section for $t$-channel and $s$-channel single-top-quark production was calculated at NLO in QCD with
\HATHOR[2.1]~\cite{Aliev:2010zk,Kant:2014oha}.
 
The $V+$jets ($V=W,Z$) and the diboson ($WW$, $WZ$, $ZZ$) production was simulated with the
\SHERPA[2.2.1] or 2.2.2 generator~\cite{Bothmann:2019yzt} depending on the process.
The simulation of $V+$jets used the NLO matrix elements for up to two partons, and leading-order (LO) matrix elements
for up to four partons, calculated with the Comix~\cite{Gleisberg:2008fv}
and \OPENLOOPS libraries.
The diboson samples, including fully leptonic final states and semileptonic final states,
where one boson decays leptonically and the other hadronically, were generated using
matrix elements at NLO accuracy in QCD for up to one additional parton
and at LO accuracy for up to three additional parton emissions; off-shell effects and Higgs boson contributions are accounted for, where appropriate.
The calculations were matched with the \SHERPA parton shower~\cite{Schumann:2007mg} using the \MEPSatNLO prescription~\cite{Hoeche:2011fd,Hoeche:2012yf,Catani:2001cc,Hoeche:2009rj},
The \NNPDF[3.0nnlo] set of PDFs was used for the matrix-element calculation, along with the dedicated set of tuned parton-shower parameters developed by the \SHERPA authors.
The $V+$jets samples were normalised to a NNLO prediction~\cite{Anastasiou:2003ds}.

The production of \ttH events was modelled using the
\POWHEGBOX[v2] generator at NLO.
The events were interfaced to \PYTHIA[8.230]. The impact of using a different parton shower and hadronisation model was evaluated by showering the nominal hard-scatter events with \HERWIG[7.04].
To assess the uncertainty associated with the NLO generator,
the nominal samples were compared with samples generated
with \MGNLO[2.6.2] at NLO in QCD using the five-flavour
scheme and the \NNPDF[2.3nlo] PDF set; these events were
interfaced to \PYTHIA[8.230].
The \ttH cross-section was calculated at NLO QCD and NLO electroweak accuracies using
\MGNLO as reported in Ref.~\cite{deFlorian:2016spz}.
 
The production of \ttV\ and \tHq\ events was modelled at NLO using the
\MGNLO[2.3.3] generator interfaced to \PYTHIA[8.210].
 
The production of \tZq events was modelled using the \MGNLO[2.3.3] generator at LO. The events were interfaced to \PYTHIA[8.210].
The \tZq total cross-section was calculated at NLO using \MGNLO[2.3.3] with the \NNPDF[3.0nlo] PDF set.


\FloatBarrier
\section{Analysis strategy}

This section presents an overview of the analysis strategy developed for the \hpdec search;
it closely follows that of similar searches performed previously by ATLAS~\cite{ATLAS:2018jqi,ATLAS:2015ncl}.
 
\subsection{Event categorisation}

This search targets the production of a charged Higgs boson via top-quark-pair decay, $t\bar{t}\rightarrow H^{\pm}bW^{\mp}b$, followed by the decays \hpdec and $W^{\mp}\rightarrow \ell\nu$, where $\ell$ denotes an electron or muon.\footnote{The small additional signal acceptance from leptonic $\tau$-lepton decays arising from the $W$-boson decay is also taken into account.}
The resulting signal event topology is characterised by four jets in the final state, three of them originating from $b$-quarks and one from a $c$-quark, which
can be effectively exploited to suppress the background. Additional jets can also be present because of initial- or final-state radiation.
 
In order to maximise the sensitivity of the search, the preselected events are categorised into different analysis
regions depending on the number of jets (4, 5 and 6) and on the number of $b$-tagged jets (3 and $\geq$4); they are summarised in Figure~\ref{fig:fit_regions}.
Events with two $b$-tagged jets are retained only if they have an additional $b$-tagged jet selected with a looser requirement: they satisfy the 70\% OP but fail the nominal (i.e. 60\%) OP (denoted \twb).
Therefore, a total of nine analysis regions are considered:
\fojtwb, \fijtwb, \sijtwb, \fojthb, \fijthb, \sijthb, \fojfob, \fijfob and \sijfob,
where ($n$j, $m$b + $k$bl) indicates $n$ selected jets, among which $m$ and $k$ are $b$-tagged with the nominal and loose $b$-tagging requirements, respectively;
event categories with at least three $b$-tagged jets passing the 60\% OP are used in the statistical analysis and thus are referred to as \enquote{fit regions}.
 
As explained in Section~\ref{subsec:rew}, the \twb regions are used to derive data-based corrections to improve the modelling of the \ttbar background.
The definition of these analysis regions ensures a background composition as close as possible to what is expected in fit regions with three $b$-tagged jets.
The main signal regions are \fojthb and \fijthb; for $\BR=0.1\%$, the signal purity in
these regions can reach a maximum value over the explored \hpm range of 2.8\% for the former and 1.9\% for the latter. The \sijthb region has lower signal purity, below 1.3\% for
$\BR=0.1\%$, and it is used mainly to constrain the background modelling uncertainties via a profile-likelihood fit (see Section~\ref{sec:result}).
Finally, the \fijfob and \sijfob regions offer a clean control sample to calibrate the \ttbarb background, while data belonging to the \fojfob region are used to recover acceptance for signal events with a $c$-quark misidentified as a $b$-jet.
 
Figure~\ref{fig:regions} shows the fractions of the different background components in the analysis regions.
In all regions, \ttbar production is the main source of SM background; it accounts for more than 80\% of the expected background.
The \ttbar background composition depends on the jet and $b$-jet multiplicities.
Analysis regions that include a \twb requirement have a background composition consisting of up to 62\% \ttbarlight and up to 28\% \ttbarb,
with an increasing fraction of \ttbarb\ at higher jet multiplicity.
The fit regions with a \thb requirement have higher fraction of \ttbarb background, up to 61\%.
Most of the \ttbarlight background events in these regions have a $b$-tagged $c$-jet from the hadronic $W$-boson
decay, in addition to the two $b$-jets from the top-quark decays.
The background in regions with the \fob requirement is mostly \ttbarb.
The \ttbarc and other backgrounds are small in all analysis regions;
the \ttbarc background mostly populates the \twb regions, accounting for up to 22\% of the expected SM background contribution.
Non-\ttbar production contributes up to 13\% in any analysis region.
 
\begin{figure*}[htbp]
\begin{center}
\includegraphics[width=0.9\textwidth]{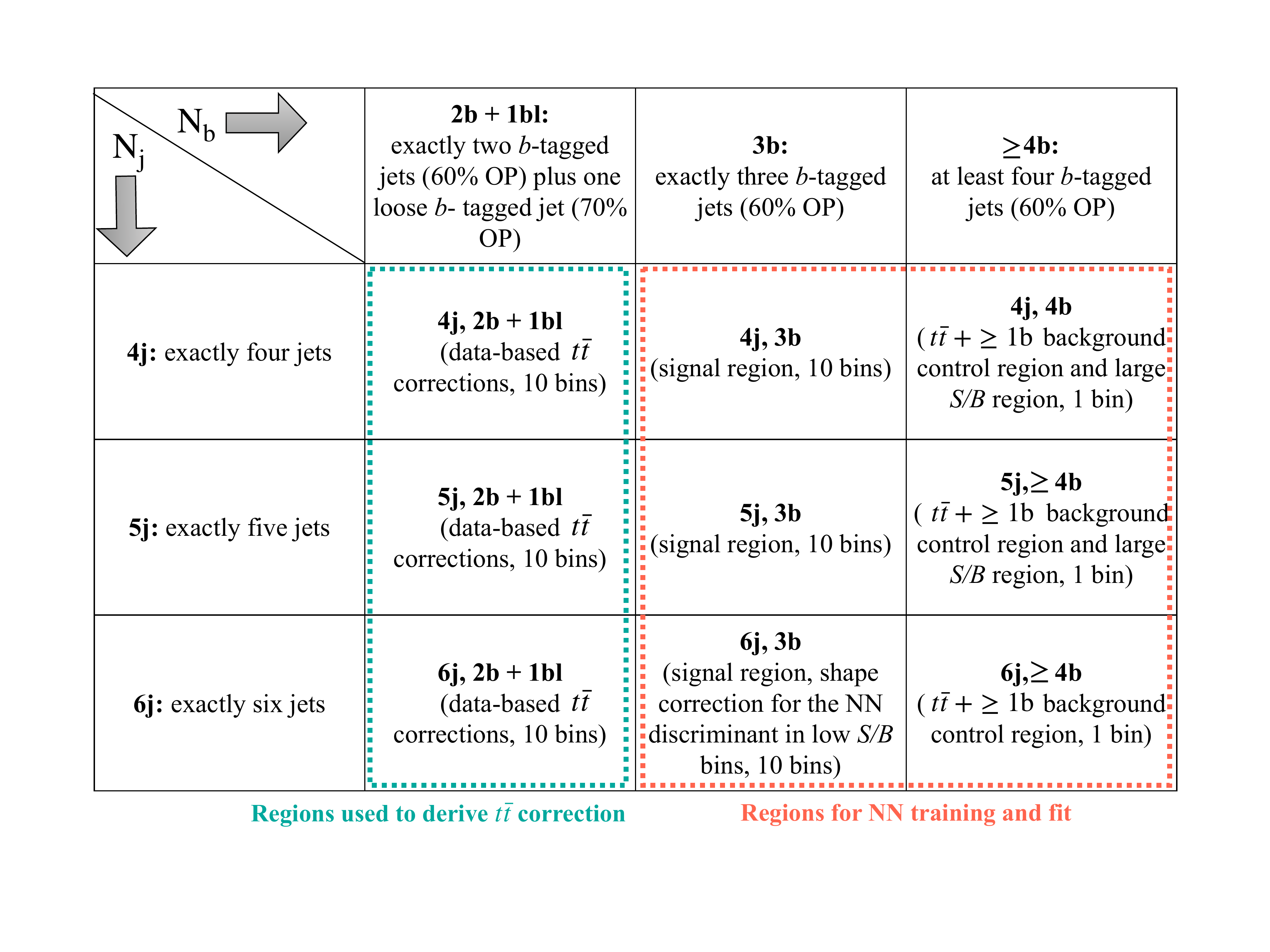}
\caption{Summary of the analysis regions along with information about their usage, as described in the text. The $b$-tagging selection is based on the DL1r algorithm's 60$\%$ efficiency OP.}
\label{fig:fit_regions}
\end{center}
\end{figure*}
 
\begin{figure*}[htbp]
\begin{center}{\includegraphics[width=0.50\textwidth]{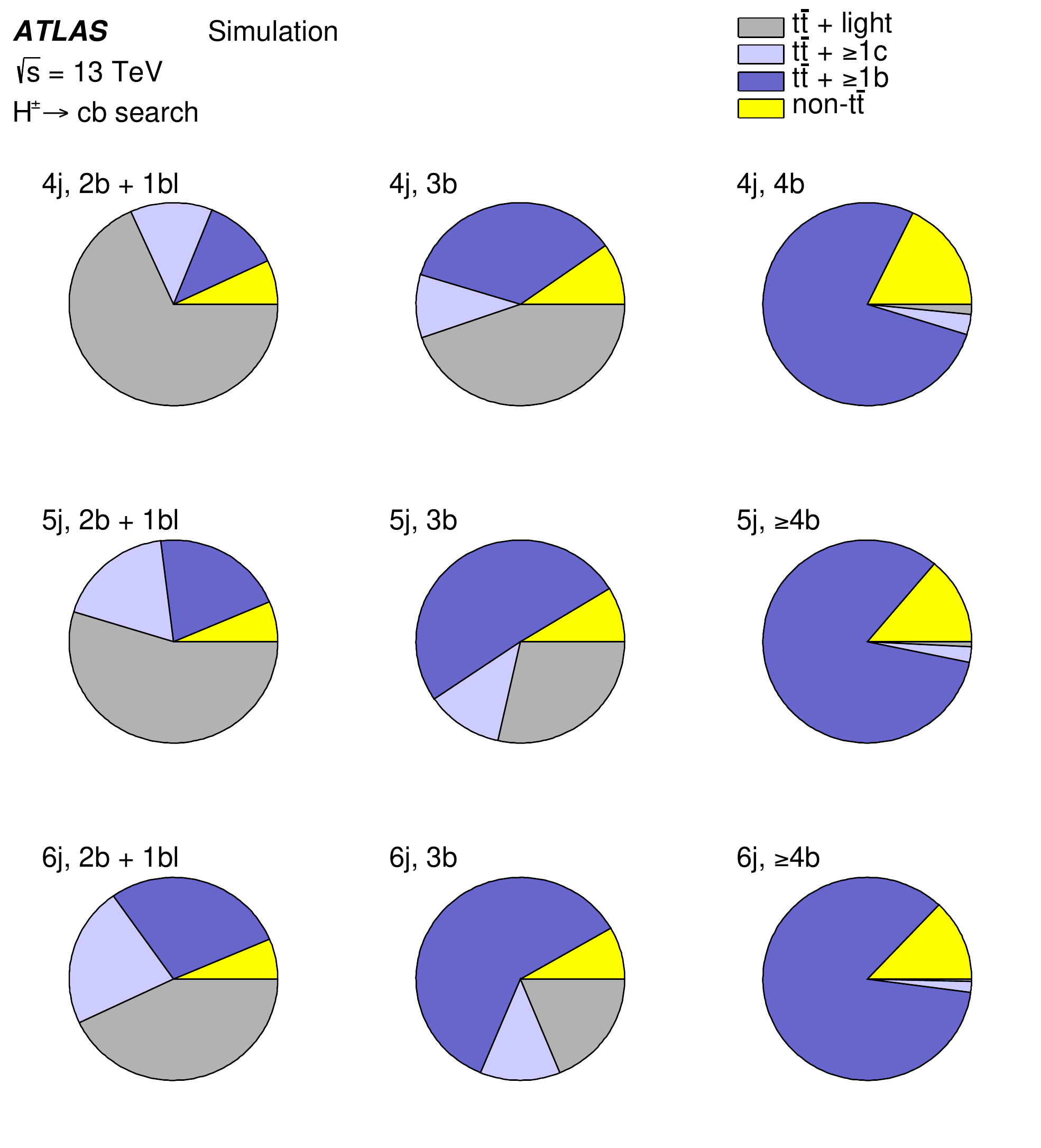}}
\caption{Fractional contributions of the various processes to the total background prediction in each analysis region.
The small contributions from $\ttbar V$, $\ttbar H$, single-top-quark, $W/Z$+jets, diboson, \tHq\ and \tZq  backgrounds are combined into a single background source referred to as ``non-$t\bar{t}$''.
The predictions for the various background contributions are obtained through the simulation as described in Section~\ref{sec:mc}. }
\label{fig:regions}
\end{center}
\end{figure*}


\subsection{\ttbar modelling}
\label{subsec:rew}

The main background for this search originates from \ttbar\ production in association with jets.
It was observed that the \ttbar\ simulation does not provide a fully satisfactory description of the jet multiplicity and transverse energy distributions
in data; this motivates the introduction of a data-based approach to correct the \ttbar simulation, similar to that
developed in recent ATLAS searches~\cite{ATLAS:2021upq,ATLAS:2021kqb}.
 
The data and the SM prediction are compared in the \twb regions separately for events with four, five or six jets.
In these three analysis regions, independent corrections for the \ttbar simulation are
derived as a function of \meff, which is defined as the scalar sum of the transverse momenta of all selected objects in the event
and \met.
 
The correction factor in a given \meff~bin~($\meffbin$)~and jet multiplicity region~(j$^{i}$) is defined as
 
\begin{equation*}
C(\meffbin, \; \text{j}^{i}) =
\frac{N^{\text{data}}(\meffbin, \; \text{j}^{i} ) - N^{\text{non-\ttbar}}( \meffbin, \; \text{j}^{i}) }{ N^{\text{\ttbar}}( \meffbin, \; \text{j}^{i}) }
\end{equation*}
 
where
$N^{\text{data}}(\meffbin,\; \text{j}^{i})$, $N^{\text{non-\ttbar}}(\meffbin,\; \text{j}^{i})$ and $N^{\text{\ttbar}}(\meffbin,\; \text{j}^{i})$ represent
respectively the yields observed in data, and the predicted non-\ttbar\ and \ttbar\ yields in the \meff\ bin and jet multiplicity under consideration.
The non-\ttbar\ yields include the small contributions from $\ttbar V$, $\ttbar H$, single-top-quark, $W/Z$+jets, diboson, \tHq\ and \tZq  backgrounds.
The fraction of subtracted non-$t\bar{t}$ background is about 7.1\% in the \fojtwb region, 6.4\% in the \fijtwb\ region, and 6.4\% in the \sijtwb\ region.
In all jet multiplicities, the derived corrections are close to unity for \meff\ above 800~\gev, and increase monotonically towards lower \meff\ values,
reaching 1.2 for $\meff=200$~\gev.
The corrections are parameterised as a function of \meff\ in each jet multiplicity bin using rational functions of varying degree.
A possible signal contamination in the \twb\ regions would mostly result in a normalisation offset for the \ttbar\ prediction, smaller than 1.5\%, due to the similarity of the $t\bar{t}$ background and signal \meff\ distribution shapes.
Such an effect is fully absorbed by the systematic uncertainties of the \ttbar\ prediction and has a numerically negligible impact on the signal extraction.
 
It was verified that after the inclusion of the data-based \ttbar corrections there is a consistently better agreement between data and the SM prediction in all analysis regions and for a wide range of observables.
Figure~\ref{fig:rew} compares the \meff distribution of data events with that of the background prediction before and after
applying the data-based correction to the \ttbar\ background simulation
in analysis regions with three $b$-tagged jets that contain events not used to derive the \ttbar\ corrections.
An improved agreement between the background prediction and data is observed, despite the fact that the correction was derived in a region with substantially lower \ttbarc and \ttbarb fractions (see Fig.~\ref{fig:regions}).
This demonstrates the appropriateness of applying these corrections to all \ttbar components (\ttbarlight, \ttbarc, and \ttbarb) in the simulation. To account for residual differences in their modelling, the associated systematic uncertainties are treated as uncorrelated between the three \ttbar components (see Sect.~\ref{subsec:ttbar_bkg_modelling}).
 
\begin{figure*}[htbp]
\begin{center}
\subfloat[]{\includegraphics[width=0.33\textwidth]{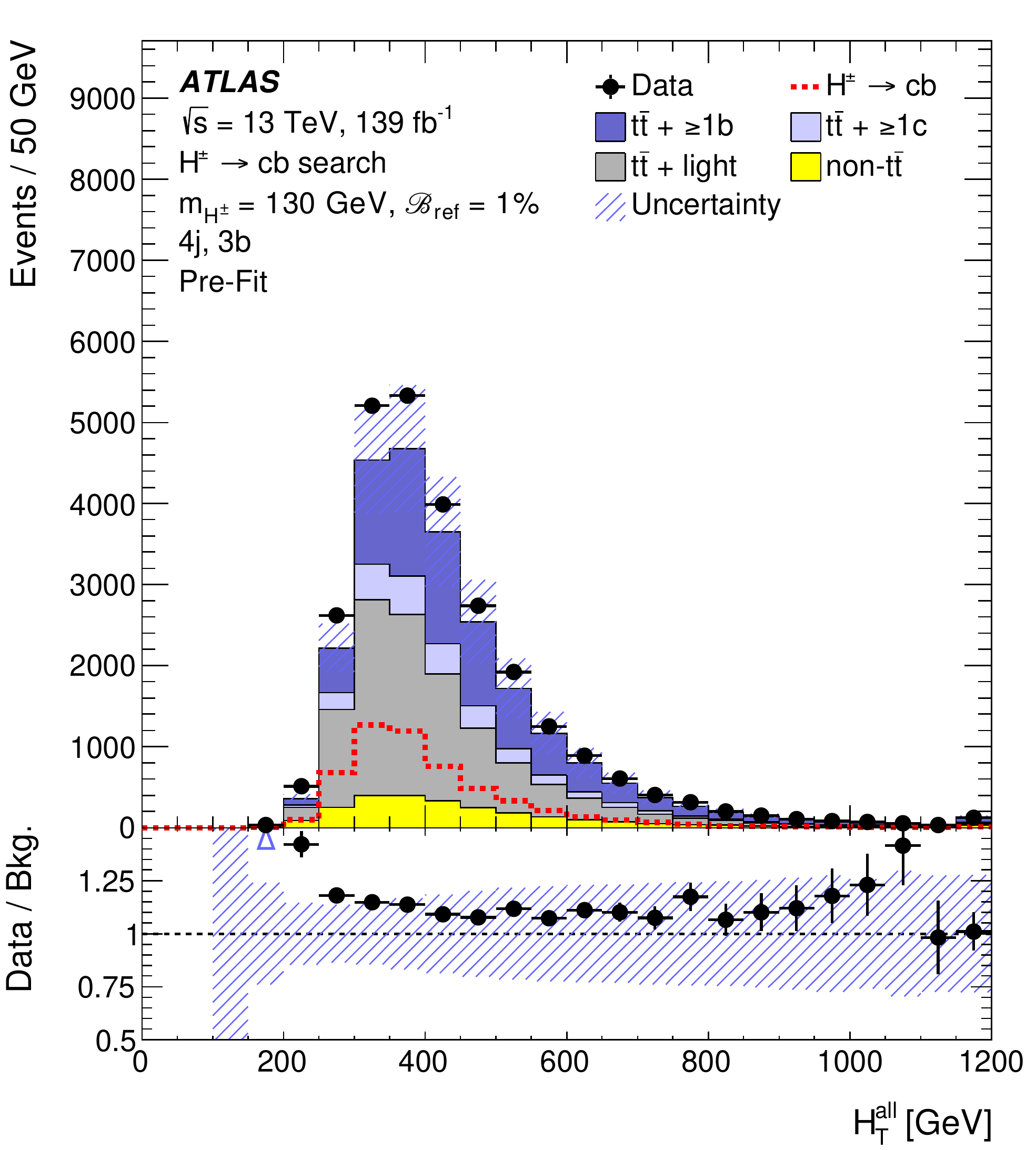}}
\subfloat[]{\includegraphics[width=0.33\textwidth]{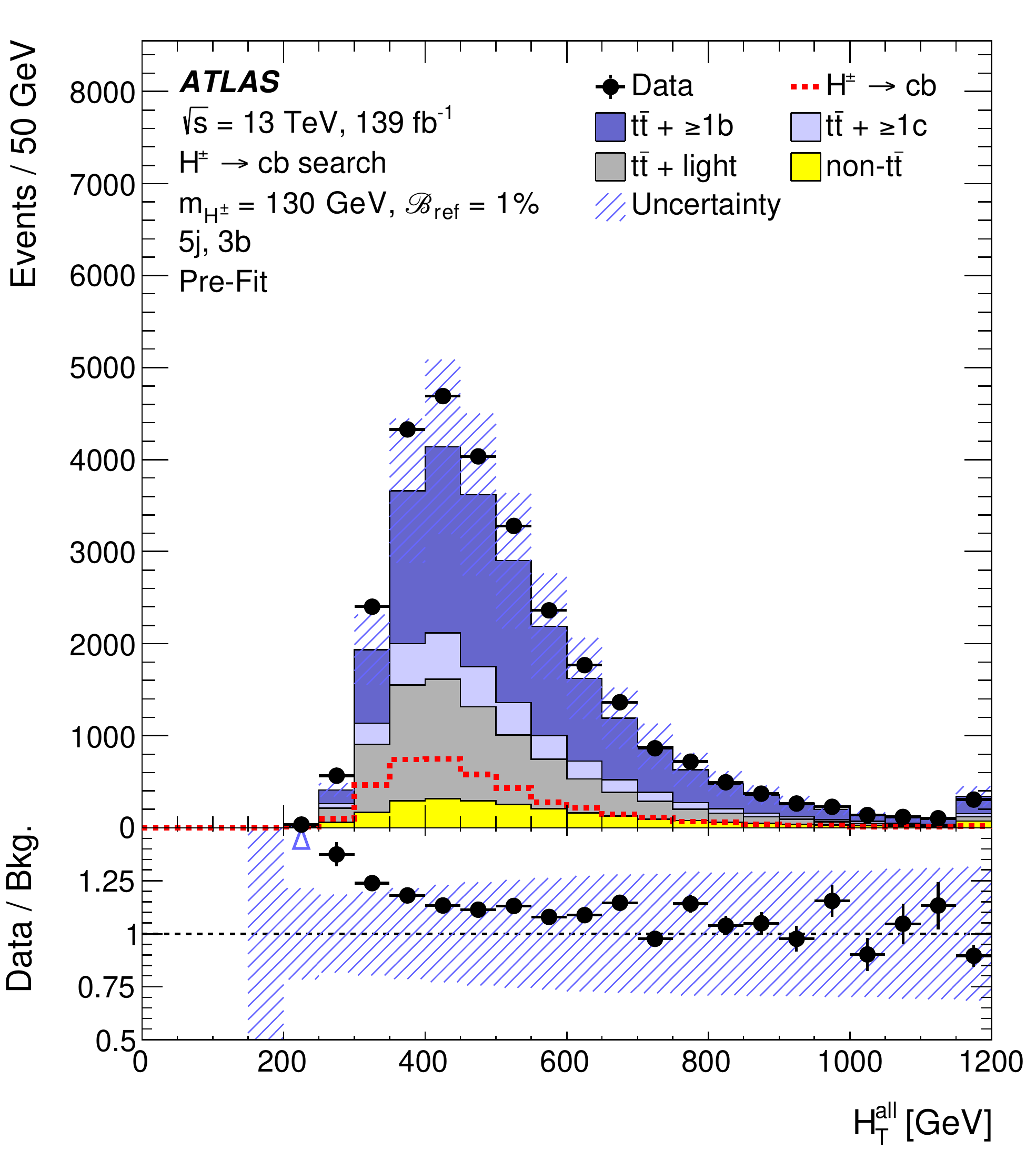}}
\subfloat[]{\includegraphics[width=0.33\textwidth]{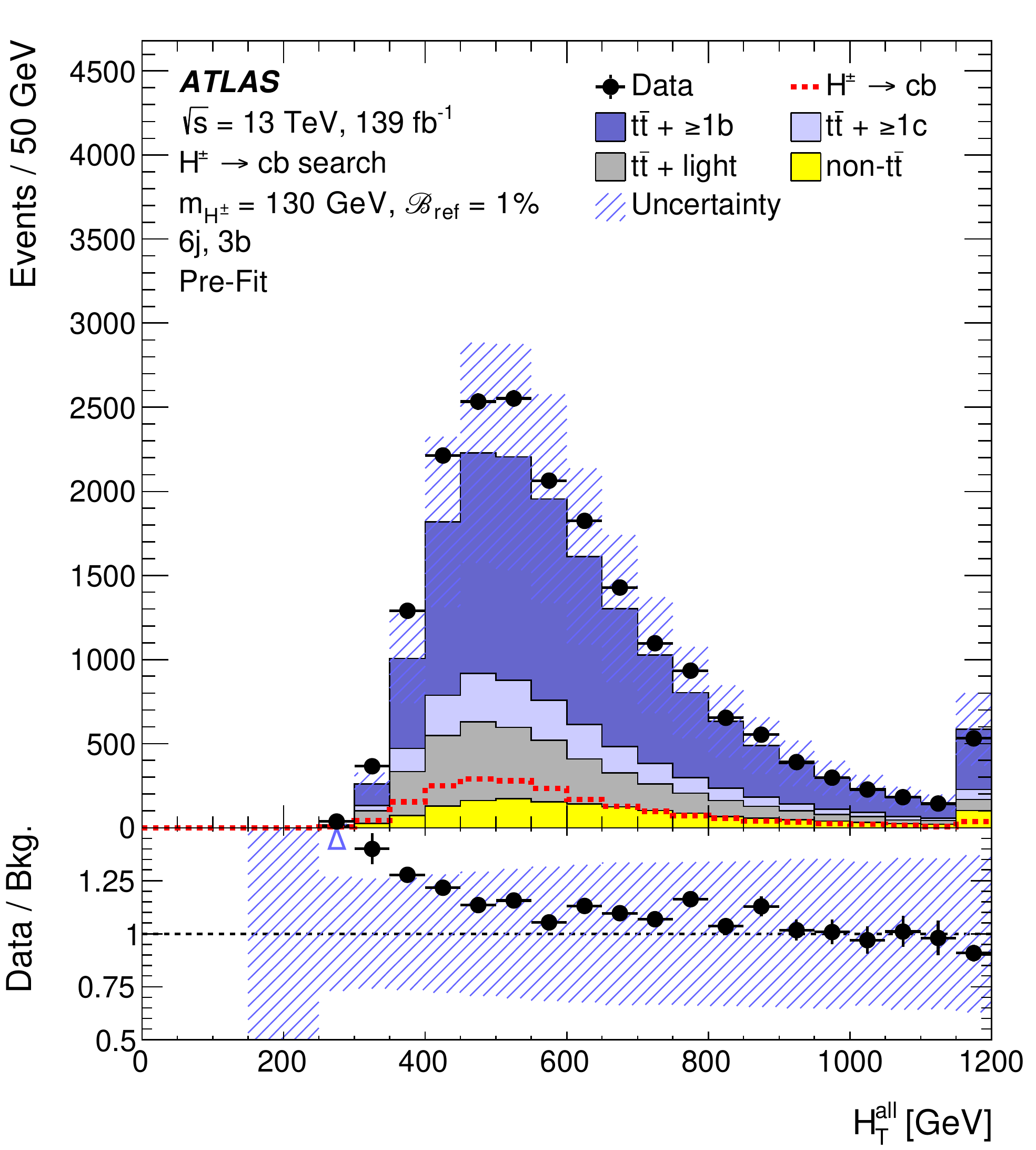}} \\
\subfloat[]{\includegraphics[width=0.33\textwidth]{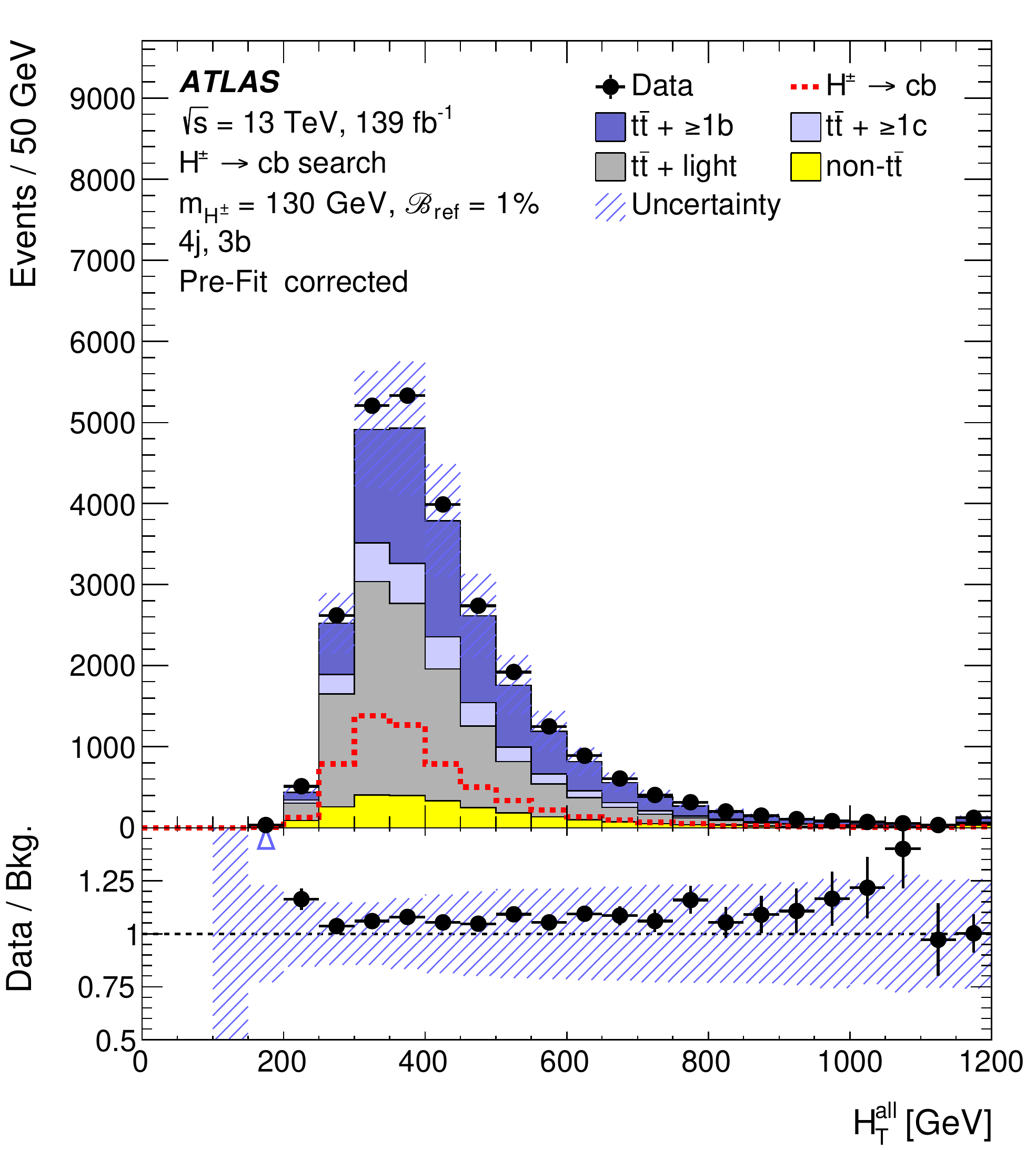}}
\subfloat[]{\includegraphics[width=0.33\textwidth]{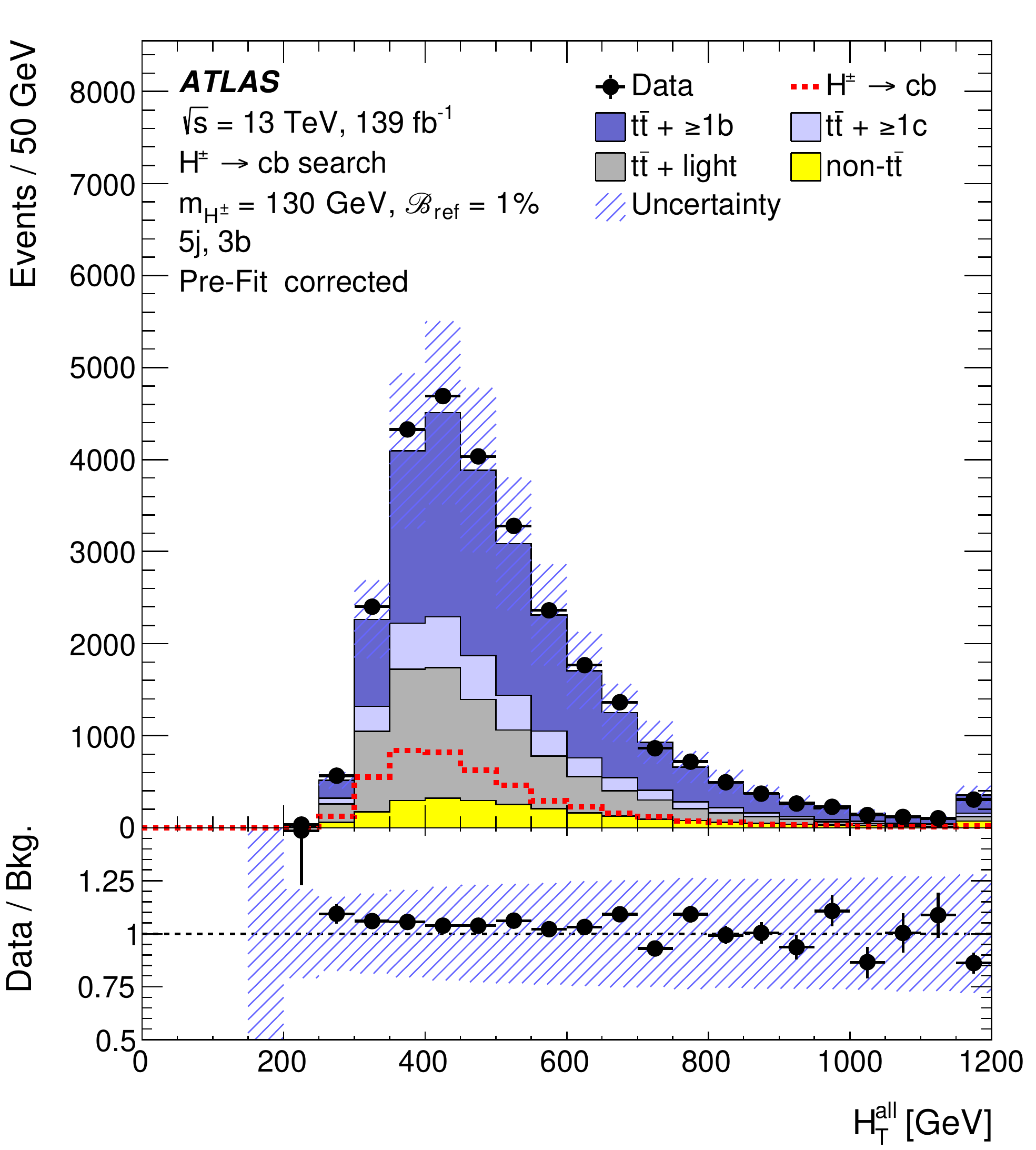}}
\subfloat[]{\includegraphics[width=0.33\textwidth]{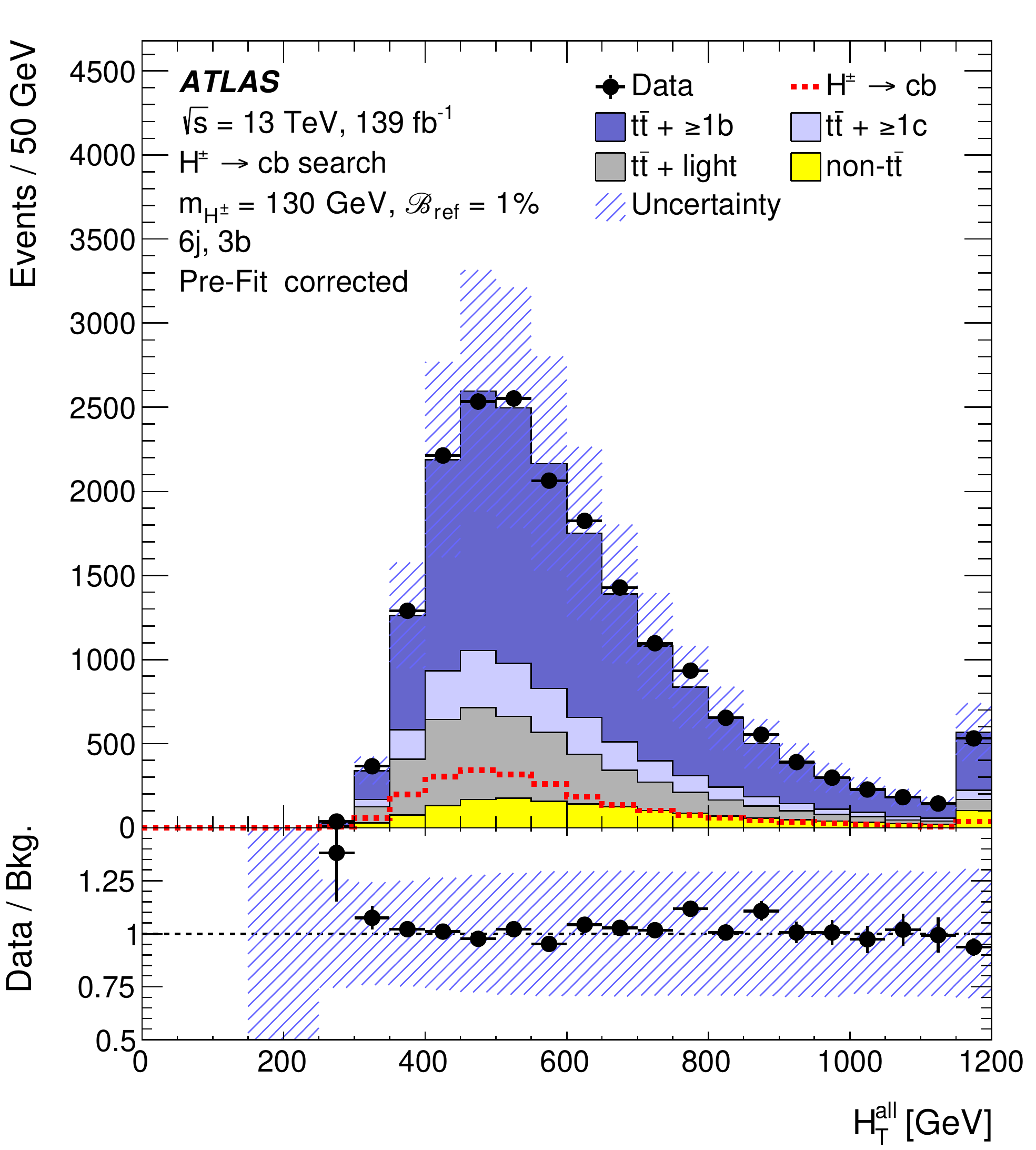}} \\
\caption{\small{
Comparison between the data and prediction for the \meff\ distribution (a-c) before and (d-f) after the inclusion of the data-based \ttbar correction (``corrected''), in regions with three $b$-tagged jets prior to the
likelihood fit to data (``Pre-Fit''), see Section~\ref{sec:result}.
The small contributions from $\ttbar V$, $\ttbar H$, single-top-quark, $W/Z$+jets, diboson, \tHq\ and \tZq  backgrounds are combined into a single background source referred to as ``non-$t\bar{t}$''.
The expected \hp signal for \hpm=~130~\gev\ is displayed as a dashed red line normalised to $\BR_{\text{ref}} =1\%$.
The bottom panels display the ratios of data to the SM background prediction (``Bkg.'') before the likelihood fit.
The hashed area represents the total uncertainty of the background. The last bin in all figures contains the overflow.
}}
\label{fig:rew}
\end{center}
\end{figure*}


\subsection{Neural network discriminant}
\label{subsec:nn}

A feed-forward neural network is used to separate the \hp signal from the large SM background.
The neural network input layer receives kinematic information about reconstructed objects,
invariant masses of jet-pair permutations, and pseudo-continuous $b$-tagging scores, for a total of 29 input variables, summarised in Table~\ref{Tab:inputsnn}.
The selected jets are initially sorted according to their pseudo-continuous $b$-tagging scores. For jets with degenerate $b$-tagging scores a $\pt$ ordering is applied.
After the jet ordering, the fourth jet in signal events is expected to originate from the charm-quark produced in the \hp decays.
The neural network uses the $\pt$, $\eta$ and $\phi$ of the first six sorted jets along with the $b$-tagging score for the fourth, fifth and sixth jets.\footnote{The input variables corresponding to the fifth (sixth) jet are set to zero in the case of events with only four (five) jets.}
These kinematic distributions and their correlations allow
the signal to be distinguished from SM background by considering the resonances produced in the event,
while the $b$-tagging scores are effective in distinguishing the three \ttbar background components (\ttbarlight, \ttbarb\ and \ttbarc) from the signal.
The $\pt$, $\eta$, and $\phi$ of the lepton, as well as \met\ and its $\phi$ angle, are also included to fully characterise the event kinematics.
Finally, three dijet invariant masses correlated with \hpm\ are obtained by calculating the invariant mass of each of the three leading jets and the fourth jet;
the inclusion of these invariant mass distributions improves the neural network's separation of signal from background by about 20\%.
 
\begin{table}
\begin{footnotesize}
\begin{center}
\caption{\small{List of input variables used in the neural network training. Jets are sorted according to their pseudo-continuous $b$-tagging scores, followed by their $\pt$, in case of having the same score.}}
\vspace{0.2cm}
\begin{tabular}{l|c}
Input variables & Number of variables\\
\hline
$\pt$, $\eta$, and $\phi$ of the first six leading jets & 18\\
$b$-tagging score of the fourth, fifth, and sixth jets & 3\\
Lepton $\pt$, $\eta$, and $\phi$ & 3\\
Missing transverse energy and its $\phi$ angle & 2\\
Invariant mass between each of the three leading jets and the fourth jet  & 3\\
\hline
Total & 29 \\
\end{tabular}
\label{Tab:inputsnn}
\end{center}
\end{footnotesize}
\end{table}
 
The neural network input layer is followed by fully connected hidden layers using rectified-linear-unit activation functions;
a sigmoid function is then used in the output layer. Batch normalisation~\cite{Ioffe:2015ovl} is applied before each hidden layer.
The training uses the Adam optimiser~\cite{Kingma:2014vow} in combination with four-fold cross-training~\cite{Kohavi95astudy},
which provides orthogonal data samples for the optimisation of the neural network hyperparameters.
The best hyperparameter configuration includes 2 layers and 190 neurons.
The impact of overtraining is mitigated with
the \enquote{dropout}~\cite{Srivastava:2014:DSW:2627435.2670313} method in combination with the \enquote{MaxNorm} constraint.
 
The neural network is trained by using events with exactly four, five or six jets, and at least three $b$-tagged jets,
corresponding to the analysis regions with highest signal purity.
The training also includes the value of the \hpm\ parameter, which for signal events is defined to be the true mass of the signal sample. In the case of  background events, a random value of the \hp\ mass, taken from the
fraction of signal masses in the input dataset, is assigned to each event~\cite{Baldi:2016fzo}. In addition to increasing the size of the training sample, the use of a mass-parameterised neural network allows the different signals to be
differentiated.
Figure~\ref{fig:nn} compares the distributions of the neural network output score (\enquote{NN score}) of the \hp signal, for $\hpm=70$~\gev\ and $\hpm=130$~\gev, and the total SM background in the three most sensitive regions, \fojthb, \fijthb\ and \sijthb.
 
\begin{figure*}[htbp]
\begin{center}
\subfloat[]{\includegraphics[width=0.33\textwidth]{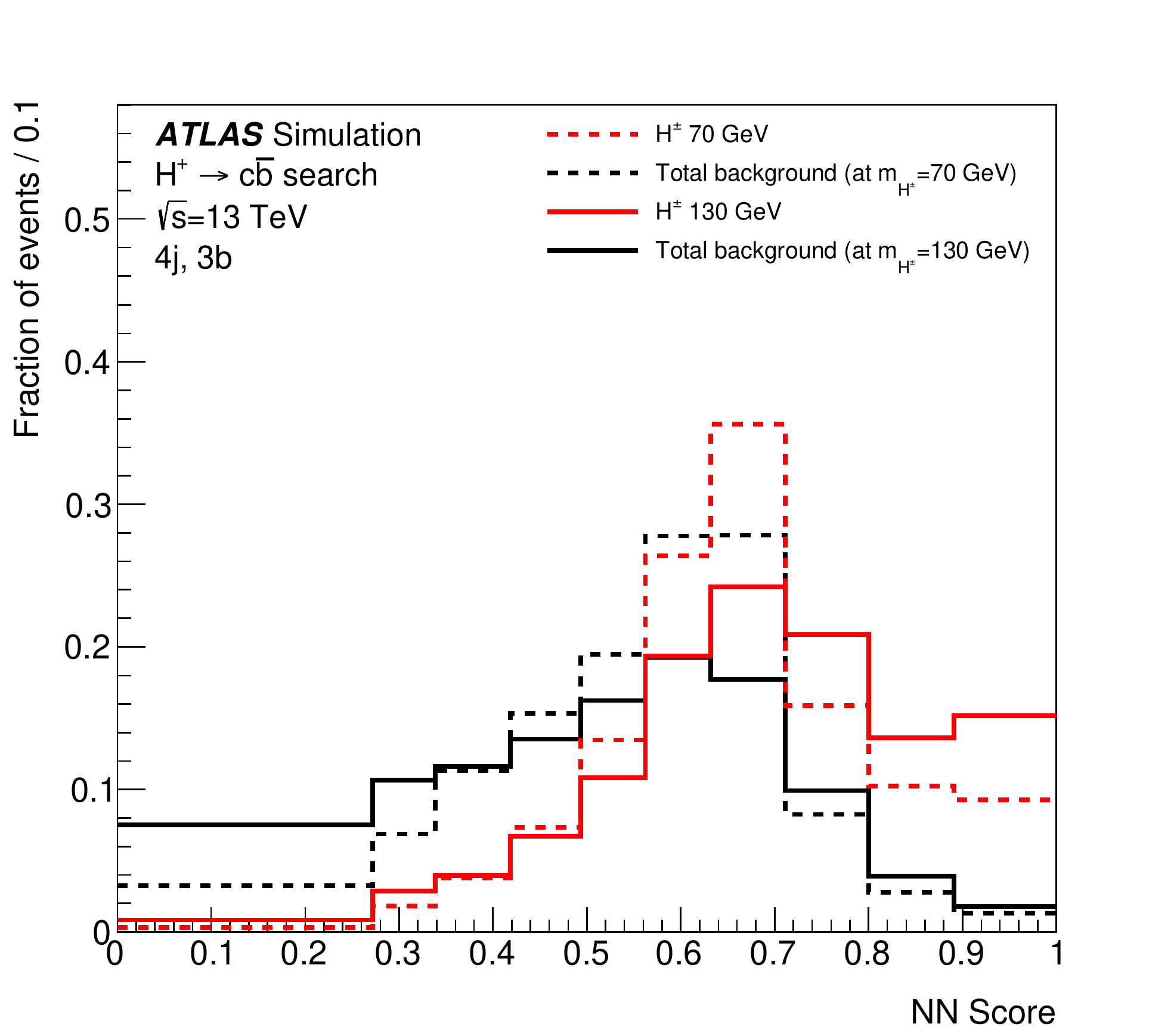}}
\subfloat[]{\includegraphics[width=0.33\textwidth]{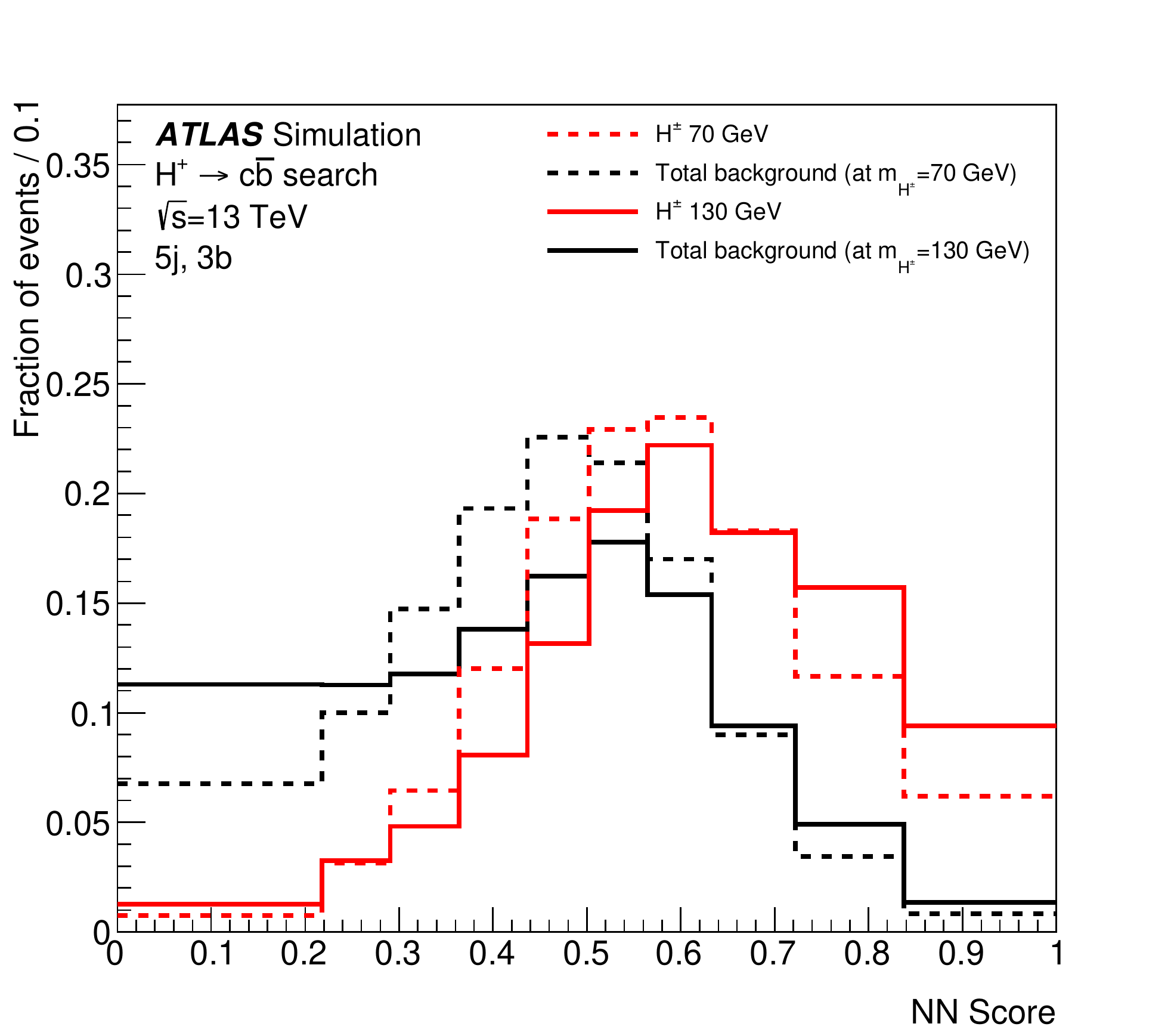}}
\subfloat[]{\includegraphics[width=0.33\textwidth]{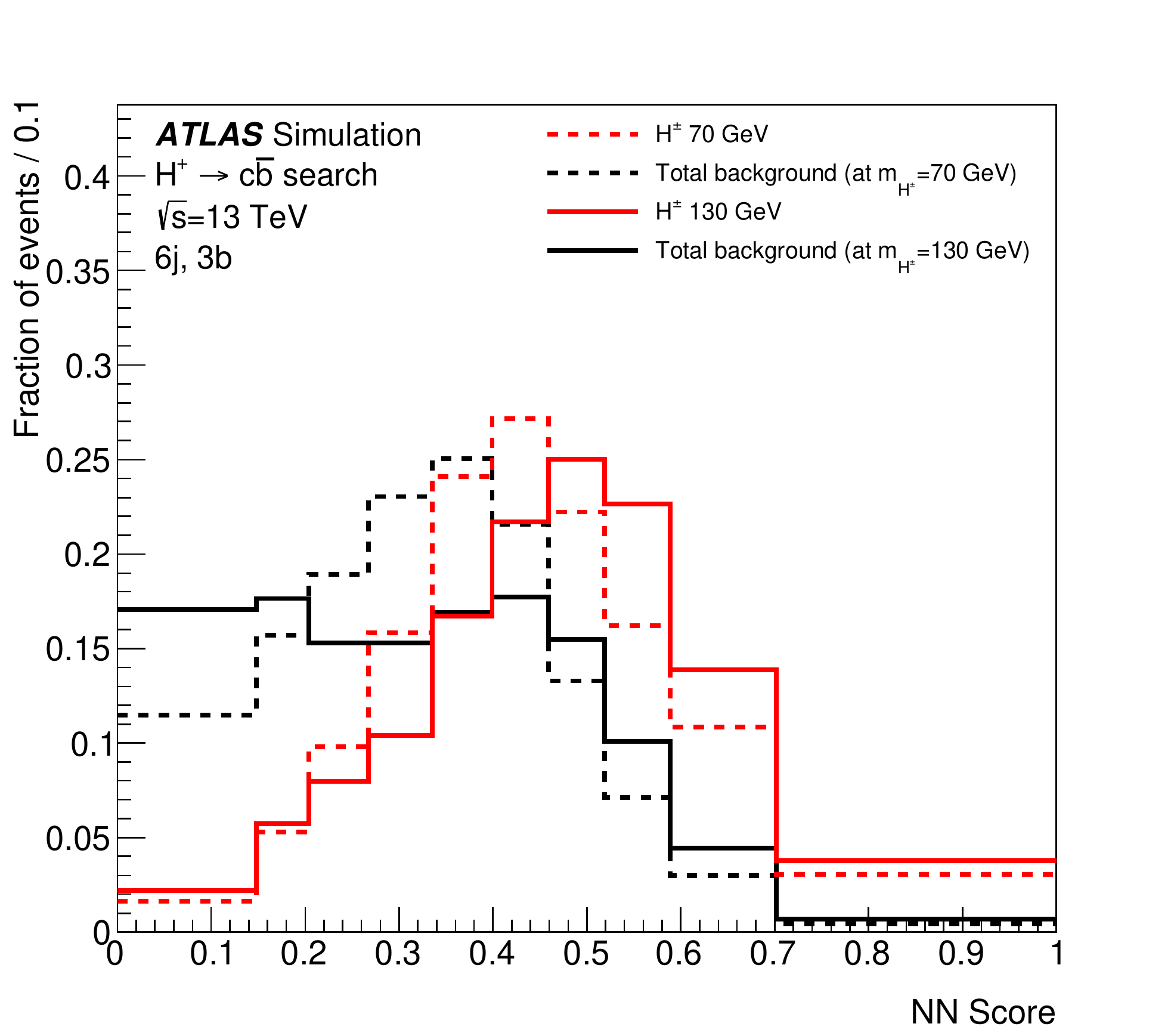}}
\caption{Comparison of the distributions of the NN score of the \hp signal with $\hpm=70$~\gev~(red dashed) and $\hpm=130$~\gev~(red solid),
and the total SM background evaluated at the same masses (black dashed and black solid, respectively) in the analysis regions (a) \fojthb, (b) \fijthb, and (c) \sijthb. Each NN score distribution is normalised to unit area.}
\label{fig:nn}
\end{center}
\end{figure*}



\FloatBarrier
\section{Systematic uncertainties}

Several sources of systematic uncertainty can affect the normalisation of the signal and backgrounds, as well as the shape of their corresponding NN scores distributions.
Each source of systematic uncertainty is considered to be uncorrelated with the other sources.
Correlations for a given systematic uncertainty are maintained
across processes and event categories, unless explicitly otherwise stated.
The following sections describe the considered systematic uncertainties.
 
\subsection{Luminosity and pile-up}
 
The uncertainty in the measurement of the integrated luminosity of the used dataset is 1.7\%~\cite{ATLAS-CONF-2019-021};
it affects the overall normalisation of all processes estimated from the simulation.
The uncertainty is derived using the LUCID-2 detector
for the baseline luminosity measurements \cite{Avoni:2018iuv}, from a calibration of the luminosity scale using $x$--$y$ beam-separation scans.
 
An uncertainty is assigned to the modelling of pile-up in simulation to account for differences between the predicted and
measured inelastic cross-sections in a given fiducial volume~\cite{ATLAS:2016ygv}.
 
The uncertainties in the luminosity measurement and pile-up modelling are treated as being correlated across the analysis regions and all physics processes.
 
\subsection{Object reconstruction}
 
Uncertainties associated with electrons and muons arise from the trigger, reconstruction,
identification and isolation efficiencies, as well as the momentum scale and resolution~\cite{ATLAS:2020auj,ATLAS:2019qmc,Aad:2020uyd,Aad:2019wsl}.
They are measured using data enriched in $Z\to \ell^+\ell^-$ and $J/\psi\to \ell^+\ell^-$ events ($\ell =e, \mu$).
Uncertainties in jet measurements arise from the jet energy scale
and resolution, and the efficiency to pass the jet-vertex-tagger requirements~\cite{ATLAS:2017bje,ATLAS:2015ull}.
The largest contribution comes from the jet energy scale, the uncertainty of which is split into 29 uncorrelated components, and depends on jet $\pt$ and $\eta$, jet flavour, pile-up treatment, and simulation of the hadronic shower shape. The jet energy scale is calibrated with a series of simulation-based corrections
and measurements in data samples enriched in photon or $Z$ boson production in association with jets or in multijet production.
 
Uncertainties associated with energy scales and resolutions of leptons and jets
are propagated to the $\met$ reconstruction. Additional uncertainties affecting the reconstruction of low-energy particles present in the event,
not associated with any leptons or jets, are measured in $Z\to \ell^+\ell^-$ data by studying the recoil of the $Z$ boson~\cite{ATLAS:2018txj}.
 
Efficiencies to tag jets in the simulation are corrected to match the efficiencies measured in data by applying $\pt$-dependent factors.
The $b$-jet efficiencies and $c$-jet mis-tagging rates are measured in a data sample enriched in $\ttbar$ events~\cite{FTAG-2018-01,FTAG-2020-08} ,
while the light-jet mis-tagging rates are measured in a multijet data sample enriched in light-jets~\cite{ATLAS:2023lwk}.
Uncertainties affecting $b$-, $c$-, and light-jet efficiencies or mis-tagging rates are decomposed into 45, 15 and 20 uncorrelated components, respectivly.
 
All uncertainties affecting object reconstruction are treated as being correlated across the analysis regions and all physics processes, including signal.
 
\subsection{\ttbar\ background modelling}
\label{subsec:ttbar_bkg_modelling}
 
The systematic uncertainties assigned to the \ttbar\ background are designed to cover potential mismodelling of this background as a function of jet and $b$-jet multiplicities.
Since the diagrams that contribute to \ttbarb, \ttbarc, and \ttbarlight processes are different,
their associated uncertainties are treated as being uncorrelated from each other, unless otherwise stated.
 
Uncertainties associated with the choice of matrix-element generator and parton shower and hadronisation models are obtained by comparing the nominal \ttbar\
sample with alternative samples described in Section~\ref{sec:mc}.
These uncertainties are evaluated in a consistent way by first putting the alternative \ttbar\ samples through the same data-based correction
procedure utilised to correct the nominal \ttbar\ sample (Section~\ref{subsec:rew}). The \ttbar\ modelling uncertainties are then evaluated by comparing
the alternative \ttbar\ samples with the nominal samples, both sets having had the data-based correction applied.
These uncertainties are also decorrelated between different jet multiplicity regions.
 
The uncertainty due to initial- and final-state radiation (ISR/FSR) was estimated by
varying the parameters of the A14 parton shower tune~\cite{ATL-PHYS-PUB-2017-007} as described in Ref.~\cite{ATLAS:2021kqb}.
Uncertainties accounting for missing higher-order QCD corrections in the matrix-element calculation are estimated by varying the renormalisation and factorisation scales
in \POWHEGBOX[v2] independently by factors of 2 and 0.5 relative to the nominal scales choice.
Uncertainties due to higher-order QCD corrections and ISR/FSR modelling are treated as being correlated between different jet multiplicity regions.
 
A normalisation uncertainty of $50\%$ is assumed separately for \ttbarb\ and \ttbarc.
For \ttbarc, the uncertainty choice is conservative given the ability to determine this background from data and the
very limited sensitivity of the final results to this choice. In the case of \ttbarb, this choice is motivated by the observed level of
disagreement between data and prediction for this background~\cite{ATLAS:2018fwl}.
 
Systematic uncertainties in the data-based \ttbar\ corrections arise from the statistical uncertainty in the parameterisation of the correction factors and
subtraction of the non-$\ttbar$ backgrounds. These uncertainties are uncorrelated between each jet multiplicity
but correlated across \ttbarb, \ttbarc and \ttbarlight background components.
 
The background originating from \ttbar\ events with a $W$ boson decaying into a charm quark and bottom quark was modelled with dedicated samples of simulated events.
Uncertainties from the NLO generator choice, as well as from the parton shower and hadronisation models, for this subset of \ttbar\ events are estimated
by comparing the nominal prediction with alternative events generated as discussed in Section~\ref{sec:mc}.
An additional cross-section uncertainty for this process is assigned by combining in quadrature a 6\% uncertainty in the inclusive $\ttbar$ production
cross-section~\cite{Czakon:2011xx} with a 3\% uncertainty in the $V_{cb}$ measurements~\cite{Zyla:2020zbs}.
 
\subsection{Signal modelling}
 
Several normalisation and shape uncertainties are taken into account for the $H^\pm$ signal. Since the signal samples were produced with the same generator and settings as the $\ttbar$ background, no alternative signal samples have been generated. Instead, the uncertainties for the \ttbarlight background associated with the choice of matrix-element generator, parton shower and hadronisation models, as well as due to ISR/FSR, are also assigned to the signal. In addition, the uncertainty in the $\ttbar$ inclusive cross section is taken into account. These uncertainties are taken to be correlated with the \ttbarlight background and uncorrelated across jet multiplicity regions. Signal modelling uncertainties have a negligible impact on the final result.
 
\subsection{Modelling of other backgrounds}
 
Uncertainties affecting the modelling of the single-top-quark background include a $+5\%$/$-4\%$ uncertainty in the total cross-section, estimated as a weighted average
of the theoretical uncertainties in $t$-, $Wt$- and $s$-channel production~\cite{Kidonakis:2011wy,Kidonakis:2010ux,Kidonakis:2010tc}.
Uncertainties associated with the choice of NLO generator and parton shower and hadronisation model are evaluated by using
alternative samples introduced in Section~\ref{sec:mc}.
The uncertainty in the ISR and FSR modelling was estimated with the same procedure deployed to evaluate the corresponding source for the \ttbar\ background.
 
Uncertainties affecting the normalisation of the $W$+jets and $Z$+jets backgrounds are estimated for the sum of both contributions (denoted $V$+jets).
Agreement between data and the total background prediction is found to be within approximately 40\%~\cite{ATLAS:2020juj}, which is taken to be the total normalisation uncertainty correlated across all $V$+jets subprocesses. An additional 25\% uncertainty is added in quadrature to the inclusive 40\% uncertainty for each additional jet beyond the fourth~\cite{Alwall:2007fs};
this procedure results in 47\% and 52\% uncertainties in regions with five or six jets, respectively.
 
Uncertainties in the diboson background normalisation include 5\% from the NLO theory cross-sections~\cite{Campbell:1999ah}.
Similarly to the $V$+jets background, an additional 25\% normalisation uncertainty is added in quadrature for each additional jet, assuming that at leading order the diboson background contributes two jets from the decay of one of the vector bosons. Therefore, the total normalisation uncertainty is 36\%, 44\%, and 50\% for events with four jets, five jets or six jets, respectively. These uncertainties are comparable to the level of disagreement found between the measured differential cross-section for $WZ$ production as a function of jet multiplicity and that predicted by the simulation~\cite{STDM-2018-03}. For both diboson and $V$+jets backgrounds, additional shape uncertainties are neglected compared to the large assigned normalisation uncertainties, which in turn have a negligible impact on the final result.
 
Modelling uncertainties for \ttH\ production were evaluated by comparing the nominal sample
with alternative samples introduced in Section~\ref{sec:mc}.
The cross-section uncertainty for \ttH\ production was estimated to be $+9\%$/$-12\%$~\cite{deFlorian:2016spz}.
 
The uncertainty in the $\ttbar V$ and \tZq cross-sections is estimated to be 60\% based on the observed level of disagreement between data and
predictions~\cite{ATLAS:2020hpj,ATLAS:2019fwo}. The uncertainty in the \tHq\ cross-sections is conservatively assumed to be 50\%.
These backgrounds have negligible impact on the results.


\FloatBarrier
\section{Results}
\label{sec:result}

\FloatBarrier

To test for the presence of a signal, a joint analysis of the NN score distributions in regions with three $b$-jets and the total yields in regions with four or more $b$-jets is performed.
The NN score is binned in ten bins in all analysis regions with three $b$-jets.
The statistical analysis uses a binned likelihood function ${\cal L}(\mu,\boldsymbol{\theta})$ constructed as
a product of Poisson probability terms over all bins considered in the search. This function depends
on the signal-strength parameter $\mu$, defined as a factor multiplying the expected yield of \hpdec signal events for $\BR_{\text{ref}}=1\%$,
and $\boldsymbol{\theta}$, a set of nuisance parameters that encode the effect of systematic uncertainties on the signal and background expectations.
All nuisance parameters are subject to Gaussian or log-normal constraints in the likelihood.
Therefore, the expected total number of events in a given bin depends on $\mu$ and $\boldsymbol{\theta}$.
 
For a given value of $\mu$, the nuisance parameters $\boldsymbol{\theta}$ allow variations of the expectations for signal and background
according to the corresponding systematic uncertainties, and their fitted values result in the deviations from
the nominal expectations that globally provide the best-fit to the data.
This procedure allows the impact of systematic uncertainties on
the search sensitivity to be reduced by taking advantage of the highly populated background-dominated bins included in the likelihood fit.
Statistical uncertainties in each bin of the predicted NN score distributions are taken into account by dedicated parameters in the fit.
The best-fit branching fraction is obtained by performing a binned likelihood fit to the data under the signal-plus-background
hypothesis, maximising the likelihood function ${\cal L}(\mu,\boldsymbol{\theta})$ over $\mu$ and $\boldsymbol{\theta}$.
 
The fitting procedure was initially validated through extensive studies using fits to real data where bins of the NN score distributions with
signal contamination above 5\% or 10\% were excluded (referred to as `blinding' or `loose blinding' requirements).
In both cases, the robustness of the model for systematic uncertainties was established by verifying the stability of the fitted background
when varying assumptions about some of the leading sources of uncertainty.
After this, the data blinding requirements
are removed and a fit under the signal-plus-background hypothesis is performed. Further checks involve a comparison of the fitted
nuisance parameters before and after removal of the blinding requirements, and their values are found to be consistent.
The fit results are also validated by comparing the NN score distributions in data with post-fit background predictions in the \twb\ regions (not directly used in the fit),
and by performing an extensive comparison between data and post-fit background predictions for several kinematic variables in all considered analysis regions. The fit is found to consistently improve the modelling of all inspected observables.
 
The test statistic $q_\mu$ is defined as the profile likelihood ratio,
$q_\mu = -2\ln({\cal L}(\mu,{\hat{\boldsymbol{\theta}}}_\mu)/{\cal L}(\hat{\mu},\hat{\boldsymbol{\theta}}))$,
where $\hat{\mu}$ and $\hat{\boldsymbol{\theta}}$ are the values of the parameters that
maximise the likelihood function (subject to the constraint $0\leq \hat{\mu} \leq \mu$), and ${\hat{\boldsymbol{\theta}}}_\mu$ are the values of the
nuisance parameters that maximise the likelihood function for a given value of $\mu$.
The test statistic $q_\mu$ is evaluated with the {\textsc RooFit} package~\cite{RooFit,RooFitManual}.
A related test statistic is used to determine whether the observed data is compatible with the background-only hypothesis (the so-called discovery test)
by setting $\mu=0$ in the profile likelihood ratio and leaving $\hat{\mu}$ unconstrained: $q_0 = -2\ln({\cal L}(0,{\hat{\boldsymbol{\theta}}}_0)/{\cal L}(\hat{\mu},\hat{\boldsymbol{\theta}}))$.
The $p$-value (referred to as $p_0$), representing the level of agreement between the data and the background-only hypothesis, is estimated by integrating
the distribution of $q_0$, based on the asymptotic formulae in Ref.~\cite{Cowan:2010js},
above the observed value of $q_0$ in the data.
Upper limits on $\mu$, and thus on
the branching ratio $\BR$, are derived by using $q_\mu$ in the CL$_{\textrm{s}}$ method~\cite{Junk:1999kv,Read:2002hq}.
For a given signal scenario, values of $\BR$ yielding CL$_{\textrm{s}} < 0.05$,
where CL$_{\textrm{s}}$ is computed using the asymptotic approximation~\cite{Cowan:2010js}, are excluded at $\geq 95\%$ CL.


\FloatBarrier

A binned likelihood fit under the signal-plus-background hypothesis
is performed on the NN score distributions in the six fit regions considered.
In the regions with exactly three $b$-tagged jets, which have the highest sensitivity, the NN score is distributed over ten bins;
in the regions with at least four $b$-tagged jets, which have a limited number of data events or small relative signal contributions,
only one bin is used.
The only unconstrained parameter of the fit is the signal strength.
Figures~\ref{fig:prefit} and~\ref{fig:postfit} show a comparison of the NN score distributions for
data and prediction in the regions with exactly three or at least four $b$-tagged jets,
respectively, before and after performing the fit to data. 
Tables \ref{Tab:smallyields130pre} and \ref{Tab:smallyields130post} display the yields before and after performing the fit to data for all the analysis regions.
 
The large number of events in the fit regions, together with their different background compositions, allows
the fit to place constraints on the combined effect of several sources of systematic uncertainty.
As a result, an improved background prediction is obtained with a significantly reduced uncertainty.
 
The regions with three $b$-tagged jets are used to constrain the leading shape uncertainties affecting the \ttbarlight\ and \ttbarb\ background predictions,
while regions with four $b$-tagged jets contribute by adjusting the normalisation of the \ttbarb\ background.
One of the main corrections applied by the fit is an increase of the \ttbarb normalisation by about a factor of $1.2\pm 0.2$
relative to the nominal prediction. This correction is in agreement with those found in previous similar ATLAS searches~\cite{ATLAS:2021upq,ATLAS:2021kqb} as well as in dedicated measurements of \ttbarb\ production~\cite{TOPQ-2017-12,CMS-TOP-18-002}.
Additionally, a few other nuisance parameters are adjusted by the fit (typically by 0.5 standard deviations or less), with the largest
effect across all performed fits being on the \ttbarc\ background normalisation, which is increased by a factor of $1.5\pm 0.5$.
Some of the largest \ttbarb\ modelling uncertainties (different parton shower and hadronisation models, and comparison to a NLO prediction in four-flavour scheme) are also significantly constrained by the fit, depending on the \hpm\ hypothesis being tested. These uncertainties are reduced to about one third of their original values, owing to the large statistics of \ttbarb\ events available in the analysis regions.
Beyond the constraints on a few individual uncertainties, the significant reduction of the total background uncertainty primarily derives from the anti-correlations among systematic uncertainties from different sources resulting from the fit.
 
The leading uncertainties affecting the signal extraction by the fit are found to be
related to the $c$-tagging calibration (up to $\Delta\BR$ $\sim$ $0.03 \times 10^{-2}$),
the calibration of light-jet mis-tagging rate (up to $\Delta\BR$ $\sim$ $0.03 \times 10^{-2}$) and
the choice of \ttbar\ NLO generator in fit regions with four jets (up to $\Delta\BR$ $\sim$ $0.05 \times 10^{-2}$).
Other uncertainties with a sizeable impact on the signal-strength measurement include \ttbarb and \ttbarc normalisation uncertainties.

\begin{figure*}[htbp]
\begin{center}
\subfloat[]{\includegraphics[width=0.33\textwidth]{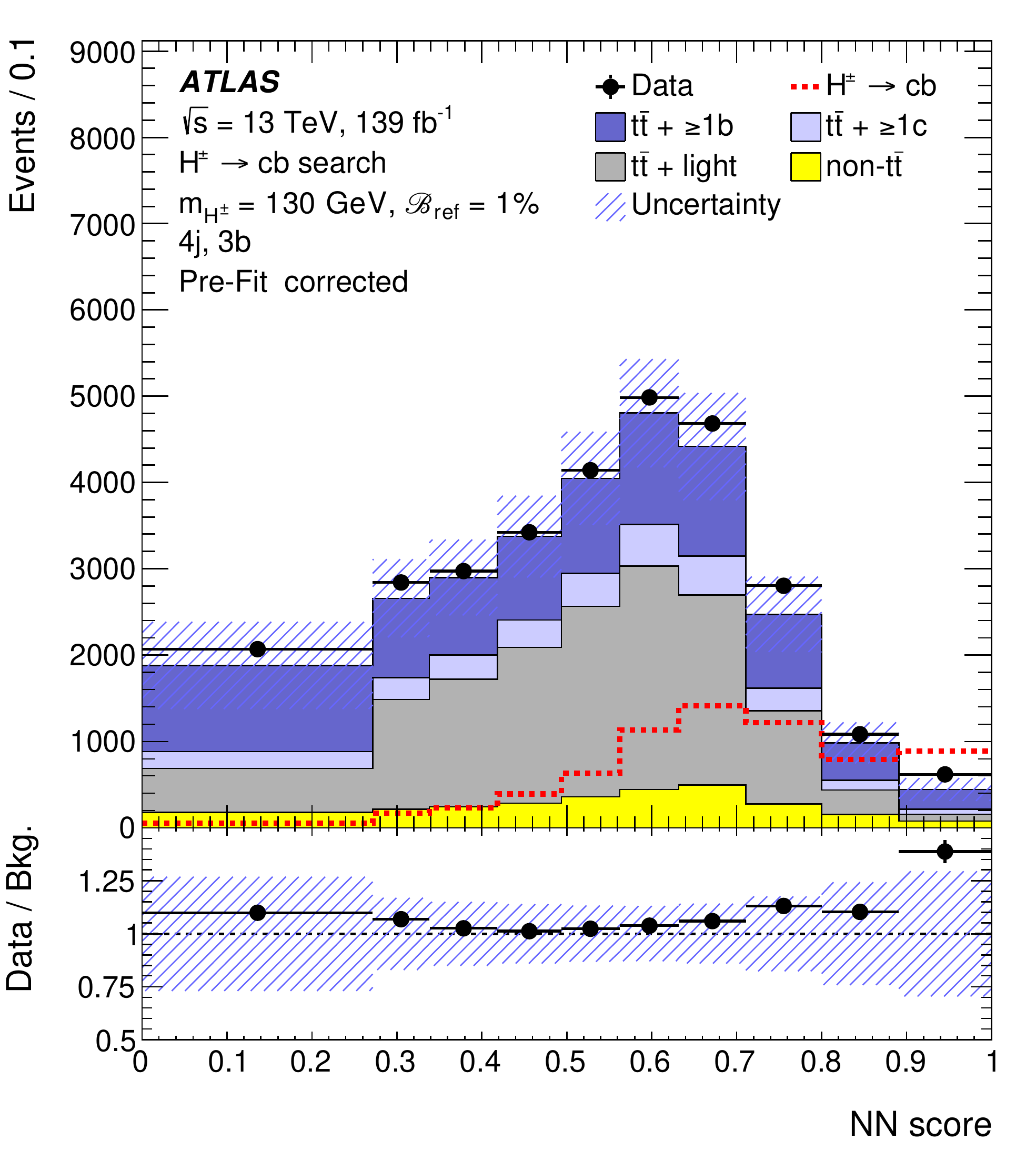}}
\subfloat[]{\includegraphics[width=0.33\textwidth]{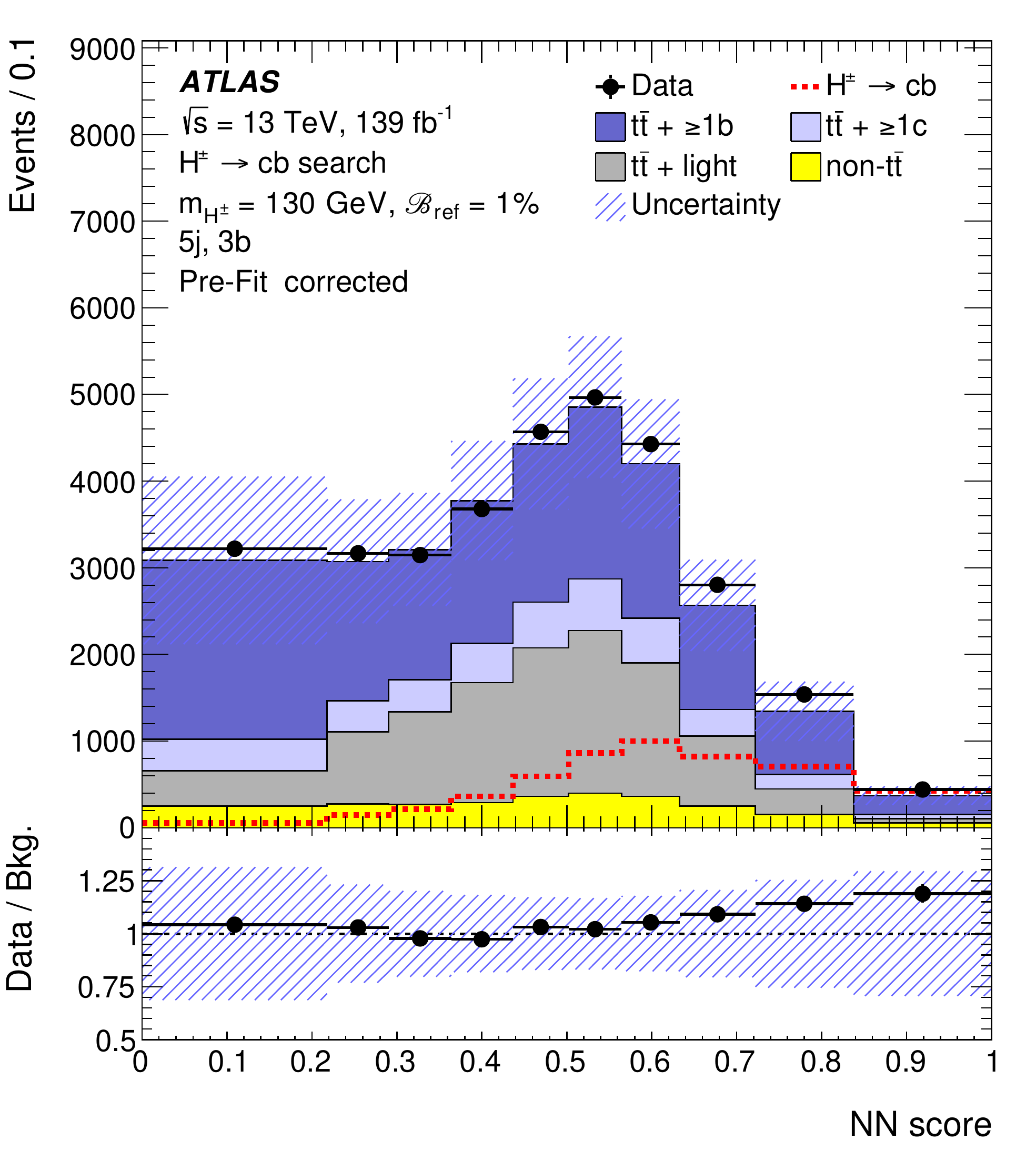}}
\subfloat[]{\includegraphics[width=0.33\textwidth]{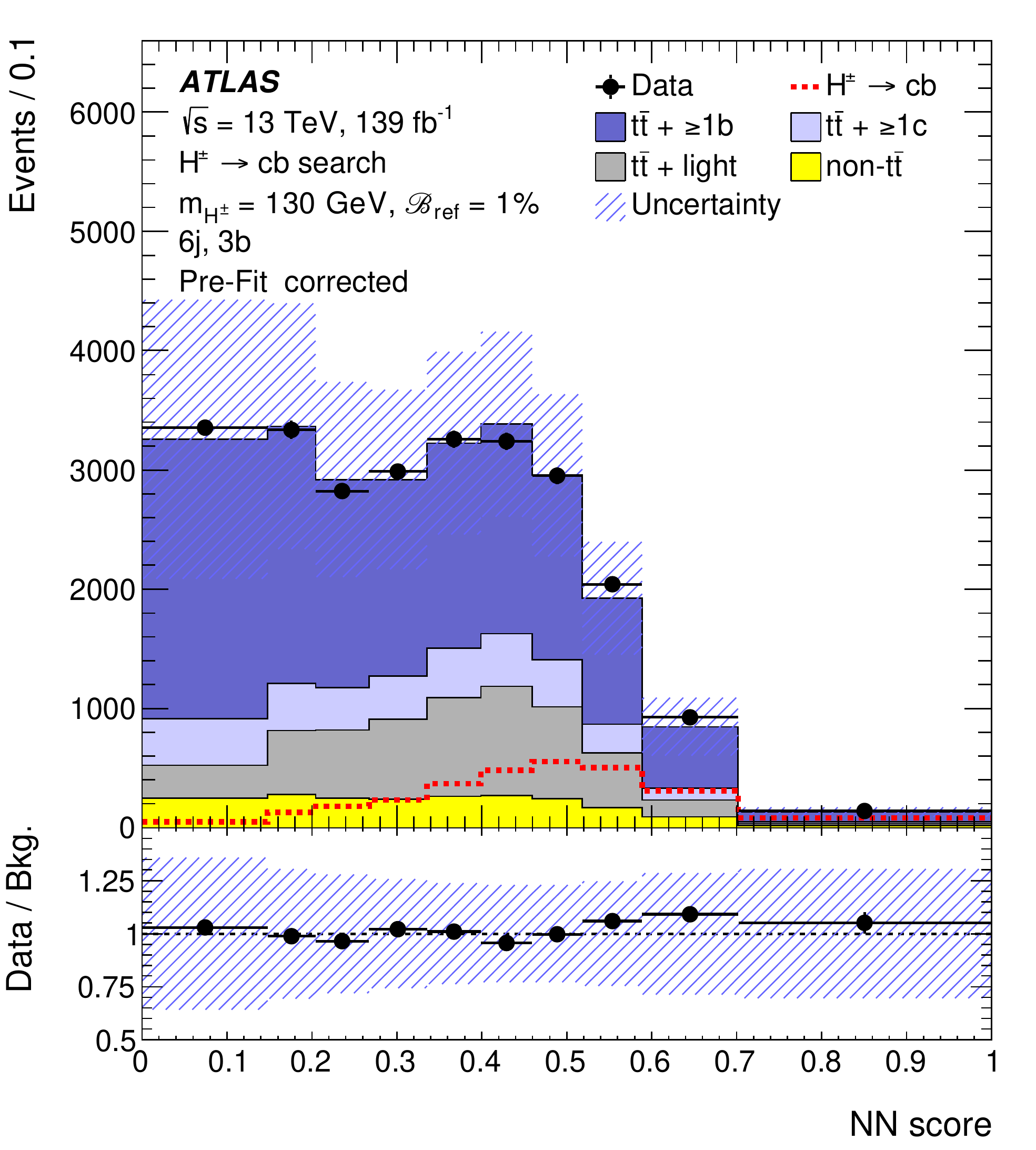}} \\
\subfloat[]{\includegraphics[width=0.33\textwidth]{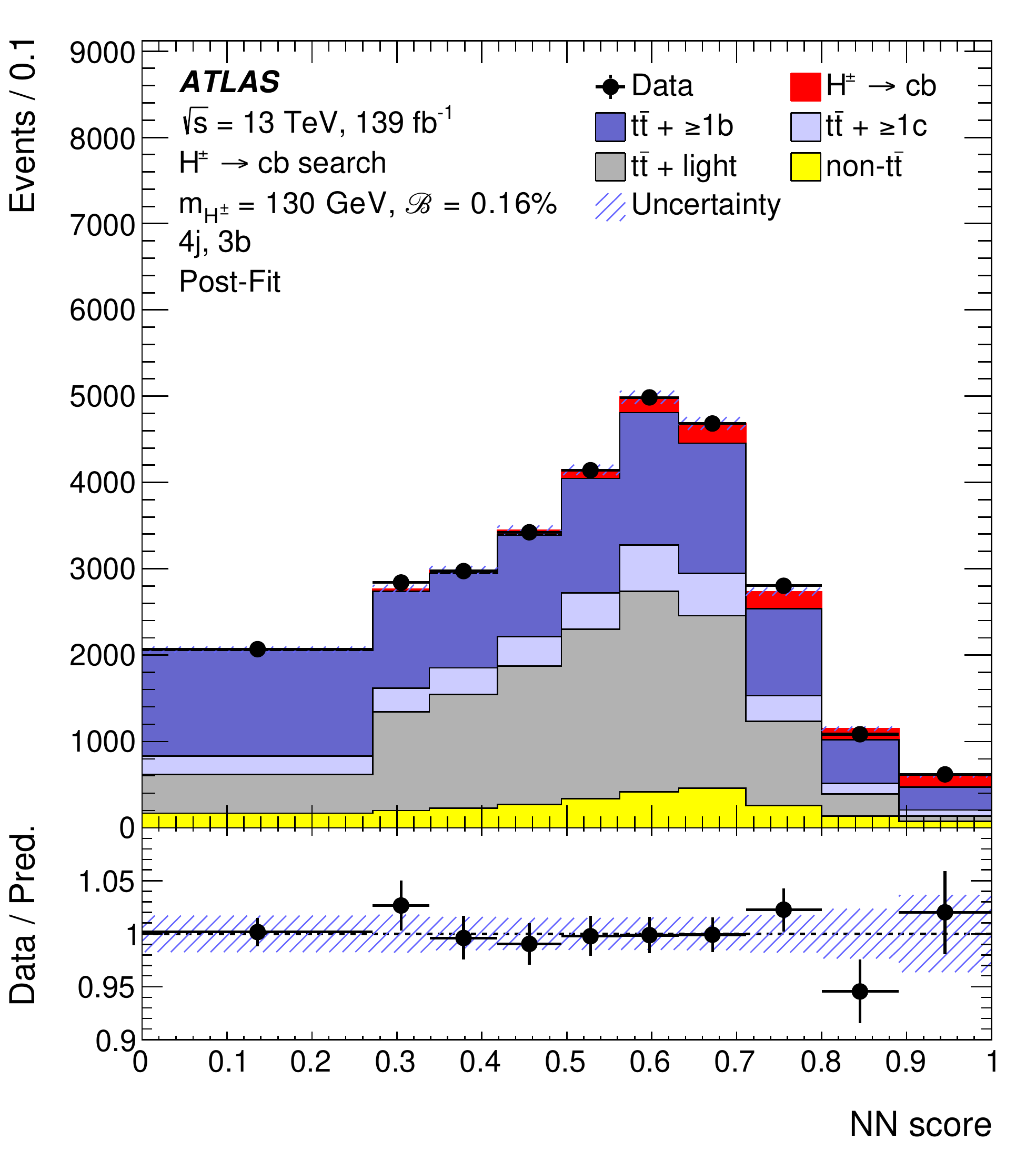}}
\subfloat[]{\includegraphics[width=0.33\textwidth]{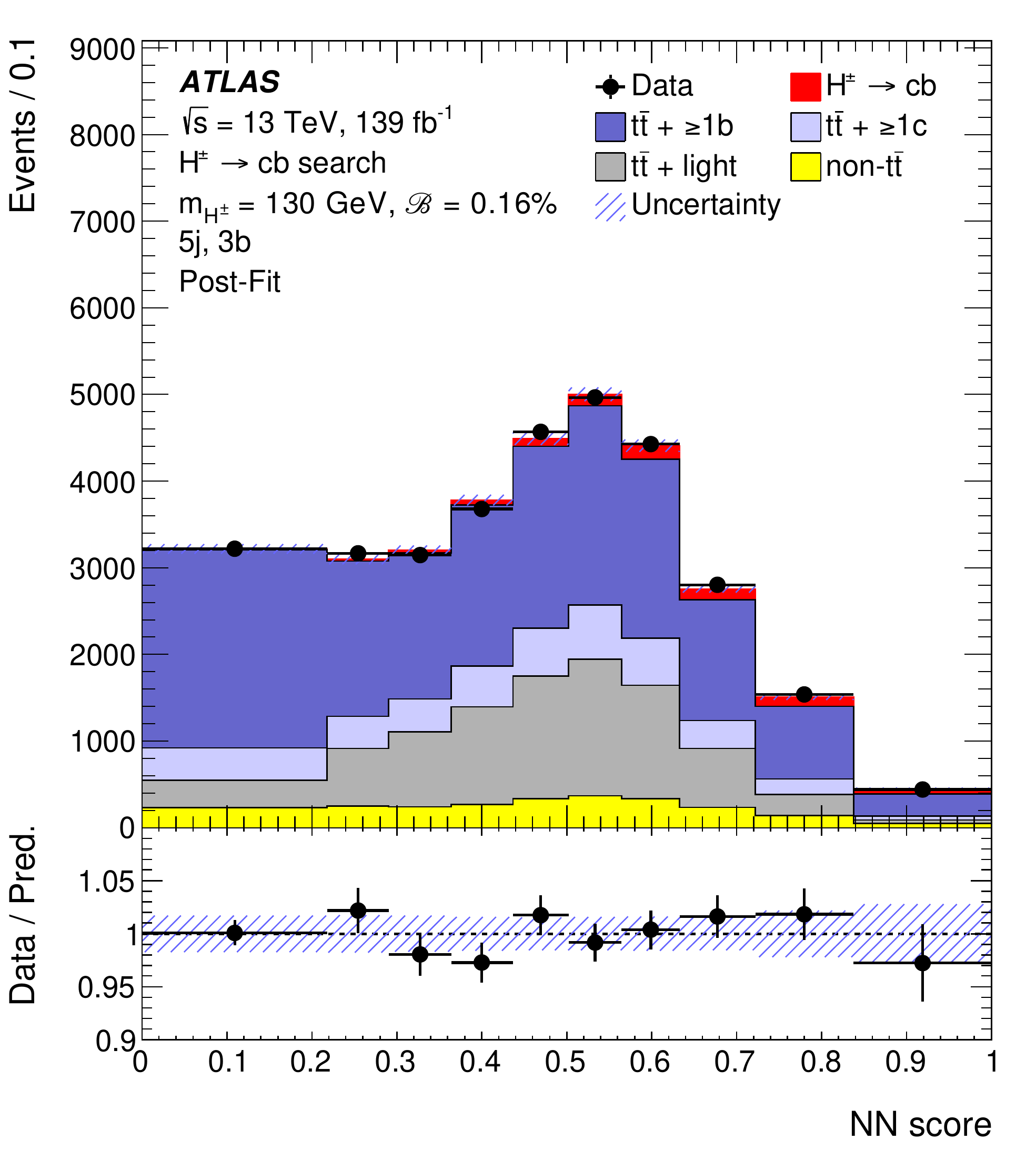}}
\subfloat[]{\includegraphics[width=0.33\textwidth]{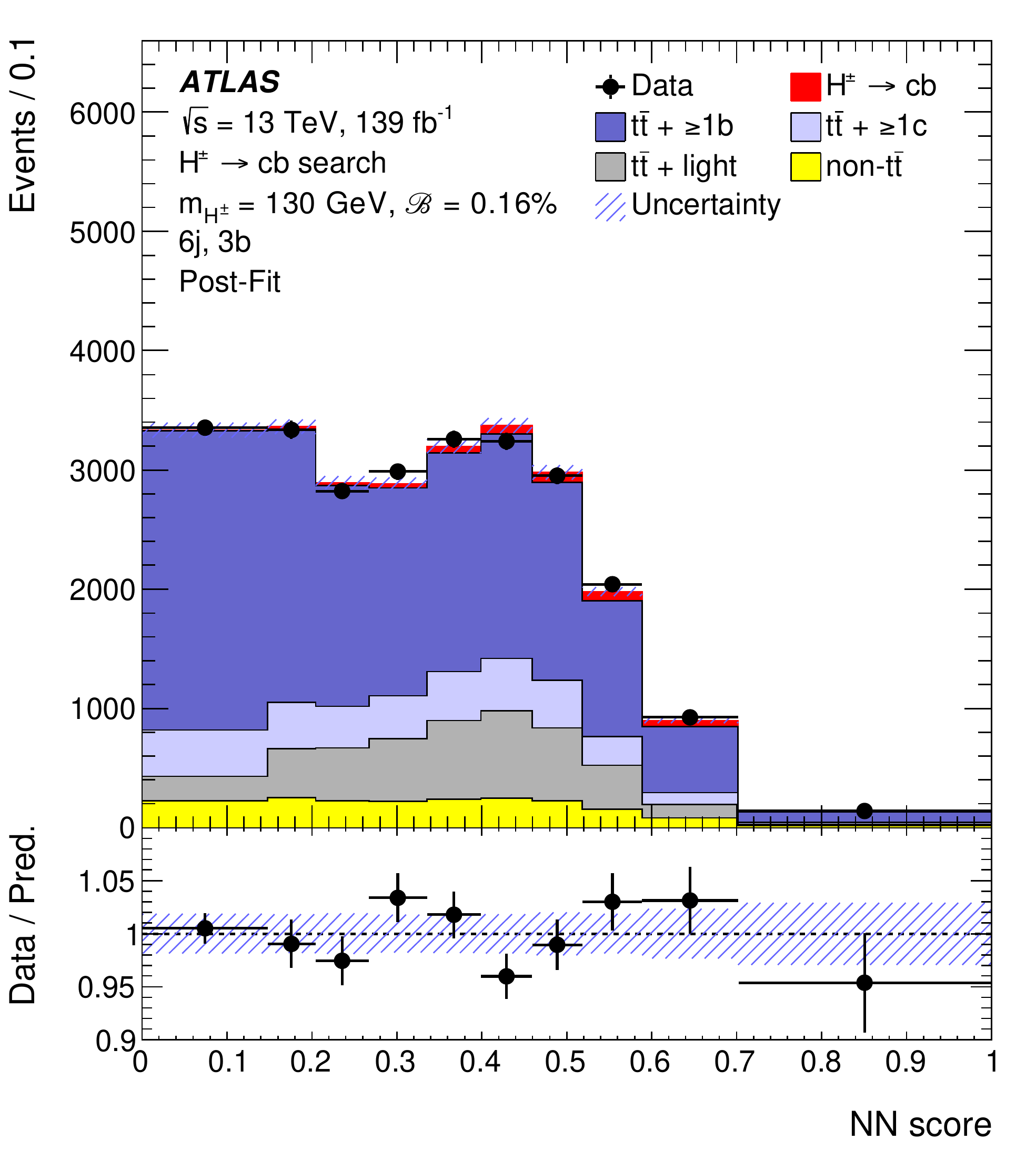}} \\
\caption{\small{Comparison between the data and prediction for the NN score distributions in the fit regions \fojthb, \fijthb\ and \sijthb\
(a-c) before the likelihood fit to data (``Pre-Fit'') and (d-f) after the likelihood fit to data (``Post-Fit'').
The small contributions from $\ttbar V$, $\ttbar H$, single-top-quark, $W/Z$+jets, diboson, \tHq\ and \tZq  backgrounds are combined into a single background source referred to as ``non-$t\bar{t}$''.
The pre-fit \hp signal for $\hpm=130$~\gev\ is displayed as a dashed red line normalised to $\BR_{\mathrm{ref}}=1\%$.
The post-fit \hp signal for $\hpm=130$~\gev\ is displayed as a red histogram normalised to the best-fit branching fraction of 0.16\%, added on top of the background prediction.
The bottom panels display the ratios of data to either the SM background prediction before the fit (``Bkg'') or the total signal-plus-background prediction after the fit (``Pred'').
The hashed area represents the total uncertainty of the background.
The \ttbar\ background prediction is corrected according to the procedure described in Section~\ref{subsec:rew}.
}}
\label{fig:prefit}
\end{center}
\end{figure*}

\begin{figure*}[htbp]
\begin{center}
\subfloat[]{\includegraphics[width=0.42\textwidth]{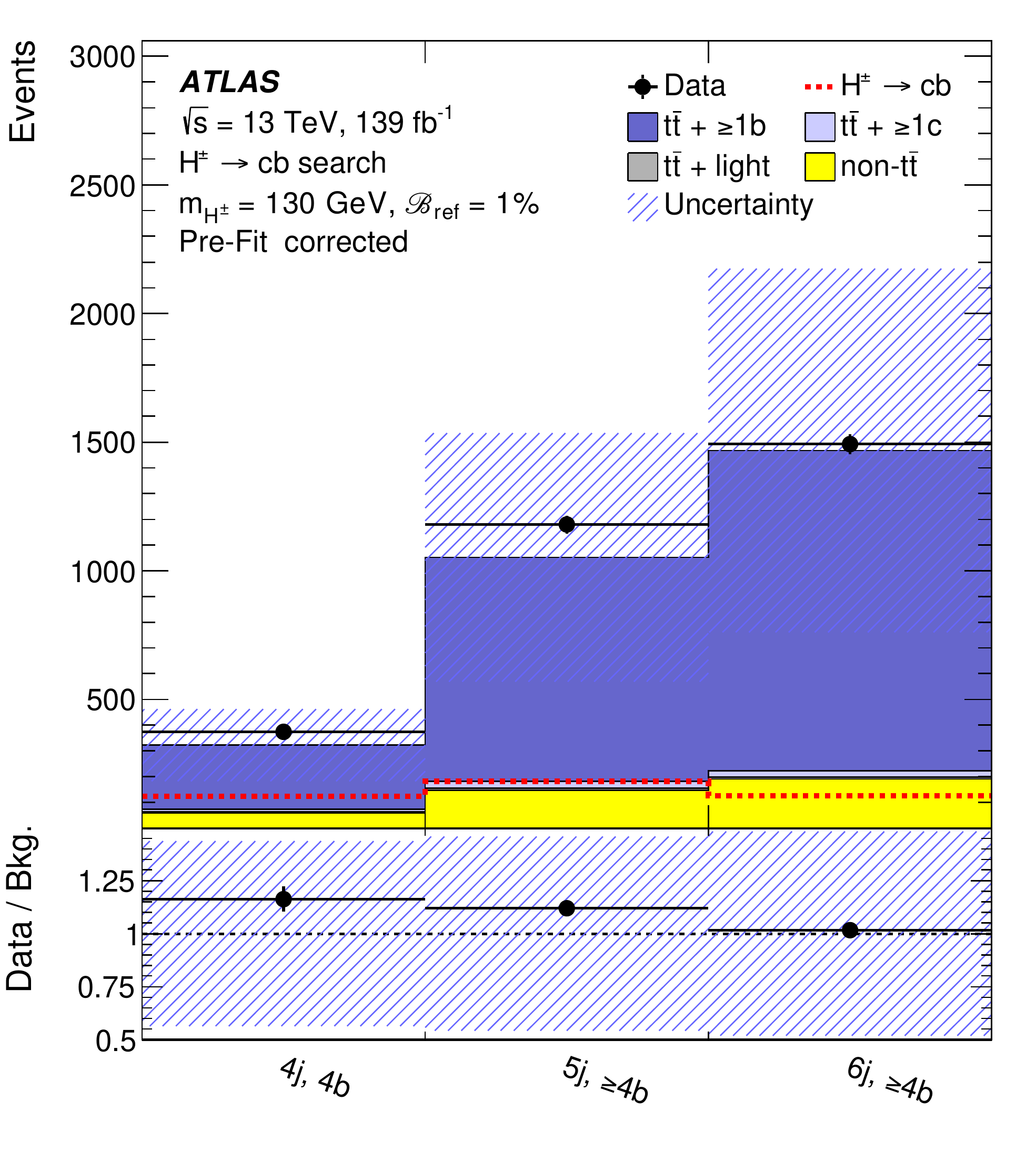}}
\subfloat[]{\includegraphics[width=0.42\textwidth]{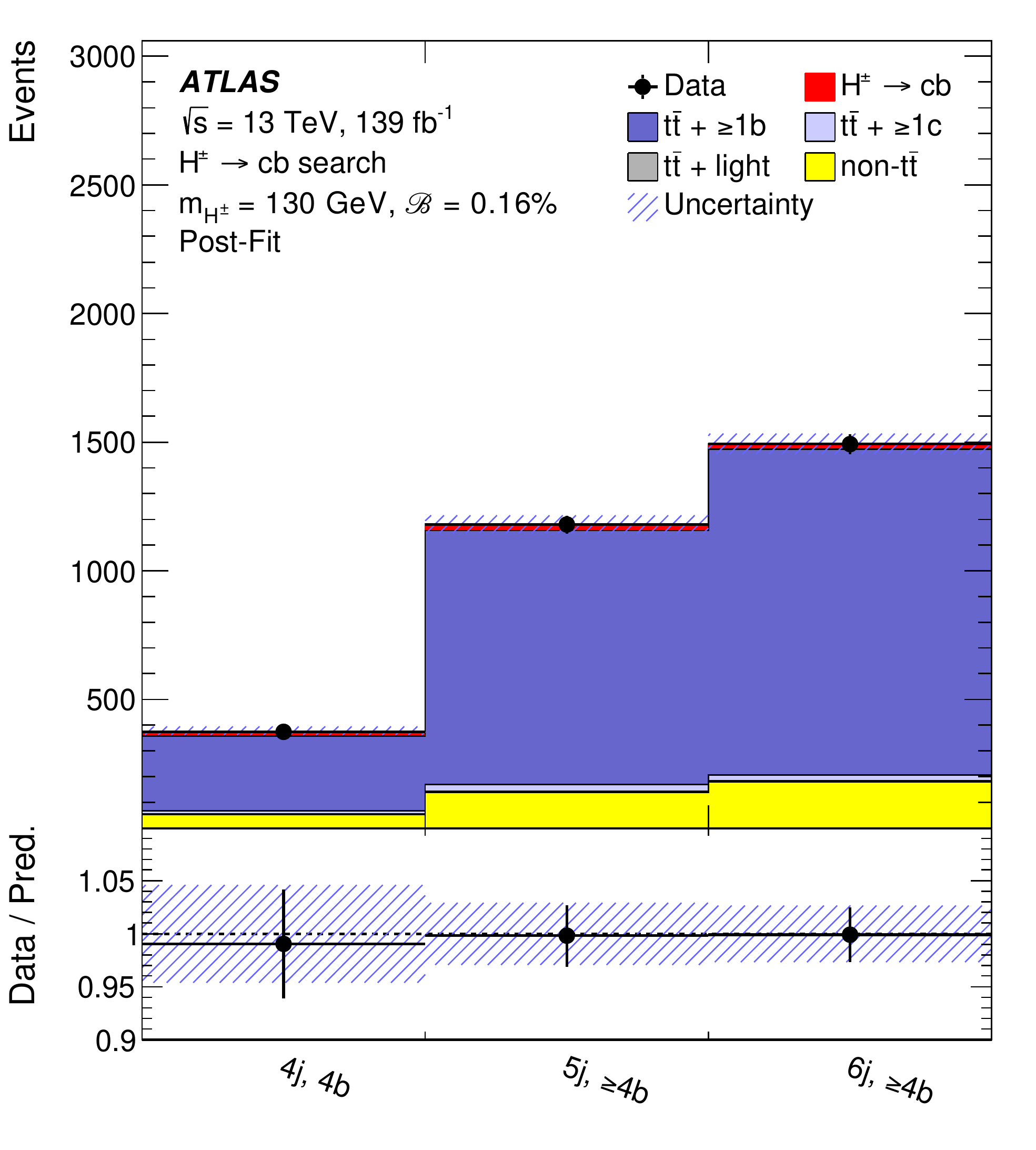}} \\
\caption{\small{Comparison between the data and prediction for the event yields in the fit regions \fojfob, \fijfob\ and \sijfob\
(a) before the likelihood fit to data (``Pre-Fit''), and (b) after the likelihood fit to data (``Post-Fit'').
The small contributions from $\ttbar V$, $\ttbar H$, single-top-quark, $W/Z$+jets, diboson, \tHq\ and \tZq  backgrounds are combined into a single background source
referred to as ``non-$t\bar{t}$''.
The pre-fit \hp signal for $\hpm=130$~\gev\ is displayed as a dashed red line normalised to $\BR_{\mathrm{ref}}=1\%$.
The post-fit \hp signal for $\hpm=130$~\gev\ is displayed as a red histogram normalised to the best-fit branching fraction of 0.16\%, added on top of the background prediction.
The bottom panels display the ratios of data to either the SM background prediction before the fit (``Bkg'') or the total signal-plus-background prediction after the fit (``Pred'').
The hashed area represents the total uncertainty of the background.
The \ttbar\ background prediction is corrected according to the procedure described in Section~\ref{subsec:rew}.
}}
\label{fig:postfit}
\end{center}
\end{figure*}
 
The best-fit $\BR$ varies between 0.06\% and 0.19\% with an absolute uncertainty ranging from about 0.13\% to 0.05\% depending on \hpm.
The total uncertainties of the measured signal strengths are dominated by systematic uncertainties.
The stability of the best-fit $\BR$ values
was confirmed by separately fitting events containing either one electron or a muon, by varying the nuisance parameters most correlated with $\BR$,
decorrelating $\BR$ across jet multiplicities and by separately fitting data collected in different data-taking years.
 
\sisetup{group-minimum-digits=4}
\begin{table}
\begin{footnotesize}
\begin{center}
\caption{\small{Pre-fit yields in each of the analysis regions considered.
The $\ttbar$ background yields are corrected with the procedure described in Section~\ref{subsec:rew}.
The contribution from $\ttbar$ production with rare $W\rightarrow cb$ decays is included in $t\bar{t}$+light.
The small contributions from $\ttbar V$, $\ttbar H$, single-top-quark, $W/Z$+jets, diboson, \tHq\ and \tZq  backgrounds are combined into a single background source referred to as ``non-$t\bar{t}$''. The \hp signal yields for $\hpm=70$~\gev\ and $\hpm=130$~\gev\ are normalised to $\BR_{\mathrm{ref}}=1\%$.
The quoted uncertainties are the sum in quadrature of statistical and systematic uncertainties of the yields,
computed taking into account correlations among processes resulting from the data-based $\ttbar$ correction procedure (see Sect.~\ref{subsec:rew}).}}
\vspace{0.2cm}
\begin{tabular}{p{5cm}|p{2cm}p{2cm}p{2cm}}
 
& 4j, 2b + 1bl  & 5j, 2b + 1bl & 6j, 2b + 1bl \\
\hline
$H^{\pm}\rightarrow cb$, $\hpm=70$~\gev  &   ~~$ \numRF{3772}{3} \pm ~~\,\numRF{269}{2}  $  &   ~~$ \numRF{2703}{3} \pm ~~\,\numRF{220}{2} $    &   ~~$ \numRF{1272}{3} \pm ~~\,\numRF{187}{2} $      \\
$H^{\pm}\rightarrow cb$, $\hpm=130$~\gev &   ~~$ \numRF{3547}{3} \pm ~~\,\numRF{265}{2}  $  &   ~~$ \numRF{2619}{3} \pm ~~\,\numRF{245}{2} $    &   ~~$ \numRF{1311}{3} \pm ~~\,\numRF{181}{2} $      \\
\hline
$t\bar{t}$+$\geq$1b                            &   ~~$ \numRF{4877}{2} \pm \numRF{2680}{2} $  &   ~~$ \numRF{7412}{2} \pm \numRF{3817}{2} $   &   ~~$ \numRF{6019}{3} \pm \numRF{3185}{2} $     \\
$t\bar{t}$+$\geq$1c                            &   ~~$ \numRF{5178}{2} \pm \numRF{2646}{2} $  &   ~~$ \numRF{6576}{2} \pm \numRF{3409}{2} $   &   ~~$ \numRF{4747}{2} \pm \numRF{2549}{2} $     \\
$t\bar{t}$+light                        &   $ \numRF{27912}{3}\pm \numRF{3596}{2} $  &   $ \numRF{19697}{3} \pm \numRF{3676}{2} $  &   ~~$ \numRF{9103}{2} \pm \numRF{2375}{2} $     \\
Non-$t\bar{t}$ backgrounds 		           &   ~~$ \numRF{2890}{3} \pm ~~\,\numRF{609}{2}  $  &   ~~$ \numRF{2298}{3} \pm ~~\,\numRF{464}{2} $    &   ~~$ \numRF{1366}{3} \pm ~~\,\numRF{305}{2} $      \\
\hline
Total background (Pre-Fit corrected)           &   $ \numRF{40858}{3} \pm \numRF{3333}{2}$  &   $ \numRF{35983}{3} \pm \numRF{3486}{2} $  &   $ \numRF{21234}{3} \pm \numRF{2994}{2} $    \\
\hline
Data  & 40\,889  & 35\,995  & 21\,210   \\
 
\multicolumn{4}{c}{ }  \\
 
& 4j, 3b & 5j, 3b  & 6j, 3b  \\
\hline
$H^{\pm}\rightarrow cb$, $\hpm=70$~\gev      &   ~~$ \numRF{6600}{3} \pm ~~\,\numRF{600}{2} $      &   ~~$ \numRF{4652}{3} \pm ~~\,\numRF{401}{2} $     &   ~~$ \numRF{2222}{3} \pm ~~\,\numRF{283}{2} $     \\
$H^{\pm}\rightarrow cb$, $\hpm=130$~\gev     &   ~~$ \numRF{5837}{3} \pm ~~\,\numRF{563}{2} $      &   ~~$ \numRF{4504}{3} \pm ~~\,\numRF{408}{2} $     &   ~~$ \numRF{2220}{3} \pm ~~\,\numRF{286}{2} $     \\
\hline
$t\bar{t}$+$\geq$1b                                &   ~~$ \numRF{8839}{2} \pm \numRF{4837}{2} $     &   $ \numRF{13888}{3} \pm \numRF{7172}{2} $   &   $ \numRF{11543}{3} \pm \numRF{6132}{2} $   \\
$t\bar{t}$+$\geq$1c                                &   ~~$ \numRF{2508}{2} \pm \numRF{1291}{2} $     &   ~~$ \numRF{3269}{2} \pm \numRF{1715}{2} $    &   ~~$ \numRF{2365}{2} \pm  \numRF{1283}{2} $   \\
$t\bar{t}$+light                            &   $ \numRF{11130}{3} \pm \numRF{1602}{2} $    &   ~~$ \numRF{7771}{2} \pm \numRF{1509}{2} $    &   ~~$ \numRF{3590}{3} \pm  ~~\,\numRF{943}{2} $    \\
Non-$t\bar{t}$ backgrounds                         &   ~~$ \numRF{2453}{3} \pm ~~\,\numRF{467}{2} $      &   ~~$ \numRF{2383}{3} \pm ~~\,\numRF{418}{2} $     &   ~~$ \numRF{1588}{3} \pm  ~~\,\numRF{308}{2} $    \\
\hline
Total background (Pre-Fit corrected)               &  $ \numRF{24931}{3} \pm \numRF{4459}{2} $     &   $ \numRF{27310}{3} \pm \numRF{6162}{2} $   &   $ \numRF{19087}{3} \pm \numRF{5419}{2} $   \\
\hline
Data  & 26\,614  & 28\,394  & 19\,302    \\
 
\multicolumn{4}{c}{ }  \\
 
& 4j, 4b  & 5j, $\geq$4b  & 6j, $\geq$4b  \\
\hline
$H^{\pm}\rightarrow cb$, $\hpm=70$~\gev       &   $ \numRF{138}{3} \pm ~~\numRF{39}{2} $     &   ~~\,$ \numRF{168}{3} \pm ~~\numRF{43}{2} $    &   ~~\,$ \numRF{119}{3} \pm ~~\numRF{56}{2} $    \\
$H^{\pm}\rightarrow cb$, $\hpm=130$~\gev      &   $ \numRF{123}{3} \pm ~~\numRF{35}{2} $     &   ~~\,$ \numRF{181}{3} \pm ~~\numRF{47}{2} $    &   ~~\,$ \numRF{124}{3} \pm ~~\numRF{59}{2} $    \\
\hline
$t\bar{t}$+$\geq$1b                               &   $ \numRF{248}{2} \pm \numRF{139}{2} $     &   ~~\,$ \numRF{871}{2} \pm \numRF{486}{2} $   &   $ \numRF{1246}{3} \pm \numRF{710}{2} $ \\
$t\bar{t}$+$\geq$1c                               &   ~~$ \numRF{10}{2}  \pm ~~~~\numRF{7}{1} $       &   ~~~~\,$ \numRF{26}{2} \pm ~~\numRF{14}{2} $     &   ~~~~\,$ \numRF{25}{2} \pm ~~\numRF{14}{2} $    \\
$t\bar{t}$+light                           &   ~~~~$ \numRF{5}{1}   \pm ~~~~\numRF{3}{1} $       &   ~~~~~~\,$ \numRF{8}{1} \pm ~~~~\numRF{6}{1} $       &   ~~~~~~\,$ \numRF{6}{1} ~~~~\pm \numRF{5}{1} $      \\
Non-$t\bar{t}$ backgrounds                        &   ~~$ \numRF{57}{2}  \pm ~~\numRF{15}{2} $      &   ~~\,$ \numRF{146}{3} \pm ~~\numRF{26}{2} $    &   $ ~~\,\numRF{191}{3} \pm ~~\numRF{41}{2} $   \\
\hline
Total background (Pre-Fit corrected)                &  $ \numRF{322}{2} \pm \numRF{142}{2} $    &   $ \numRF{1052}{3} \pm \numRF{484}{2} $  &   $ \numRF{1468}{3} \pm \numRF{708}{2} $  \\
\hline
Data  & 374  & 1\,179 & 1\,492  \\
 
\end{tabular}
 
\label{Tab:smallyields130pre}
\end{center}
\end{footnotesize}
\end{table}

\begin{table}
\begin{footnotesize}
\begin{center}
\caption{\small{Post-fit yields in each of the fit regions considered;
the analysis regions used to derive the data-based $t\bar{t}$ corrections are not used in the fit,
so their yields are not displayed.
The contribution from $\ttbar$ production with rare $W\rightarrow cb$ decays is included in $t\bar{t}$+light.
The total prediction is shown after the fit to data under the signal-plus-background hypothesis assuming \hp signal with $\hpm=130$~\gev.
The predicted yields for the \hp signal with $\hpm=70$~\gev\ are also shown for reference. The best fit-values of $\BR$ for \hp signal with $\hpm=130$~\gev\ and $\hpm=70$~\gev\ are 0.16\% and 0.07\% respectively.
The small contributions from $\ttbar V$, $\ttbar H$, single-top-quark, $W/Z$+jets, diboson, \tHq\ and \tZq  backgrounds are combined into a single background source referred to as ``non-$t\bar{t}$''.
The quoted uncertainties are the sum in quadrature of statistical and systematic uncertainties of the yields,
computed taking into account correlations among nuisance parameters and among processes.}}
\vspace{0.2cm}
\begin{tabular}{p{5cm}|p{2cm}p{2cm}p{2cm}}

& 4j, 3b & 5j, 3b  & 6j, 3b  \\
\hline
$H^{\pm}\rightarrow cb$, $\hpm=70$~\gev   &    ~~~~\,$ \numRF{450}{2} \pm ~~\,\numRF{382}{2} $   &    ~~~~\,$ \numRF{325}{2} \pm  ~~\,\numRF{275}{2} $   &  ~~~~\,$ \numRF{152}{2} \pm \numRF{130}{2} $        \\
$H^{\pm}\rightarrow cb$, $\hpm=130$~\gev  &    ~~~~\,$ \numRF{923}{2} \pm ~~\,\numRF{345}{2} $   &    ~~~~\,$ \numRF{682}{2} \pm  ~~\,\numRF{255}{2} $   &  ~~~~\,$ \numRF{333}{2} \pm \numRF{127}{2} $        \\
\hline
$t\bar{t}$+$\geq$1b                               &    $ \numRF{10678}{4} \pm ~~\,\numRF{727}{2} $   &   $ \numRF{15745}{4} \pm ~~\,\numRF{908}{2} $   &   $ \numRF{12344}{4} \pm \numRF{655}{2} $      \\
$t\bar{t}$+$\geq$1c                               &    ~~$ \numRF{2758}{2} \pm \numRF{1113}{2} $   &   ~~$ \numRF{3397}{2} \pm \numRF{1361}{2} $   &    $ ~~\numRF{2353}{3} \pm \numRF{946}{2} $       \\
$t\bar{t}$+light                           &    ~~$ \numRF{9969}{3} \pm ~~\,\numRF{694}{2} $    &   ~~$ \numRF{6378}{3} \pm ~~\,\numRF{699}{2} $    &   $ ~~\numRF{2813}{3} \pm \numRF{411}{2} $       \\
Non-$t\bar{t}$ backgrounds                        &    ~~$ \numRF{2287}{3} \pm ~~\,\numRF{413}{2} $    &   ~~$ \numRF{2195}{3} \pm ~~\,\numRF{388}{2} $     &  ~~$ \numRF{1459}{3} \pm \numRF{285}{2} $       \\
\hline
Total prediction                                  &    $ \numRF{26615}{4} \pm ~~\,\numRF{249}{2} $   &    $ \numRF{28397}{4} \pm ~~\,\numRF{313}{2} $   &   $ \numRF{19302}{4} \pm \numRF{246}{2} $     \\
\hline
Data                                              &    26\,614               &    28\,394               &    19\,302                 \\
 
\multicolumn{4}{c}{ }  \\
 
& 4j, 4b  & 5j, $\geq$4b  & 6j, $\geq$4b  \\
\hline
$H^{\pm}\rightarrow cb$, $\hpm=70$~\gev   &   ~~~~$ \numRF{8}{1} \pm ~~\numRF{7}{1} $   &   ~~~~\,$ \numRF{12}{2} \pm \numRF{11}{2} $   &    ~~~~~~\,$ \numRF{7}{1} \pm ~~\numRF{6}{1} $    \\
$H^{\pm}\rightarrow cb$, $\hpm=130$~\gev      &   ~~$ \numRF{20}{2} \pm ~~\numRF{8}{1} $    &   ~~~~\,$ \numRF{25}{2} \pm \numRF{11}{2} $   &    ~~~~\,$ \numRF{21}{2} \pm \numRF{11}{2} $     \\
\hline
$t\bar{t}$+$\geq$1b                               &    $ \numRF{291}{3} \pm \numRF{23}{2} $    &   $ ~~\,\numRF{987}{3} \pm \numRF{46}{2} $   &   $ \numRF{1266}{4} \pm \numRF{57}{2} $   \\
$t\bar{t}$+$\geq$1c                               &    ~~$ \numRF{11}{2} \pm ~~\numRF{7}{1} $      &   ~~~~\,$ \numRF{26}{2} \pm \numRF{11}{2} $    &   $ ~~~~\,\numRF{24}{2} \pm \numRF{11}{2} $     \\
$t\bar{t}$+light                           &    ~~~~$ \numRF{3}{1} \pm ~~\numRF{2}{1} $       &   ~~~~~~\,$ \numRF{4}{1} \pm ~~\numRF{4}{1} $      &   ~~~~~~\,$ \numRF{4}{1} \pm ~~\numRF{3}{1} $     \\
Non-$t\bar{t}$ backgrounds                        &    ~~$ \numRF{52}{2} \pm \numRF{13}{2} $     &   $ ~~\,\numRF{138}{3} \pm \numRF{26}{2} $   &   $ ~~\,\numRF{179}{3} \pm \numRF{40}{2} $    \\
\hline
Total prediction                                    &   $ \numRF{378}{3} \pm \numRF{17}{2} $    &   $ \numRF{1182}{4} \pm \numRF{34}{2} $  &   $ \numRF{1494}{3} \pm \numRF{38}{2} $    \\
\hline
Data                                                &   374               &   1\,179             &   1\,492               \\
\end{tabular}
\label{Tab:smallyields130post}
\end{center}
\end{footnotesize}
\end{table}


\FloatBarrier

\clearpage
\begin{figure*}[t!]
\begin{center}{\includegraphics[width=0.65\textwidth]{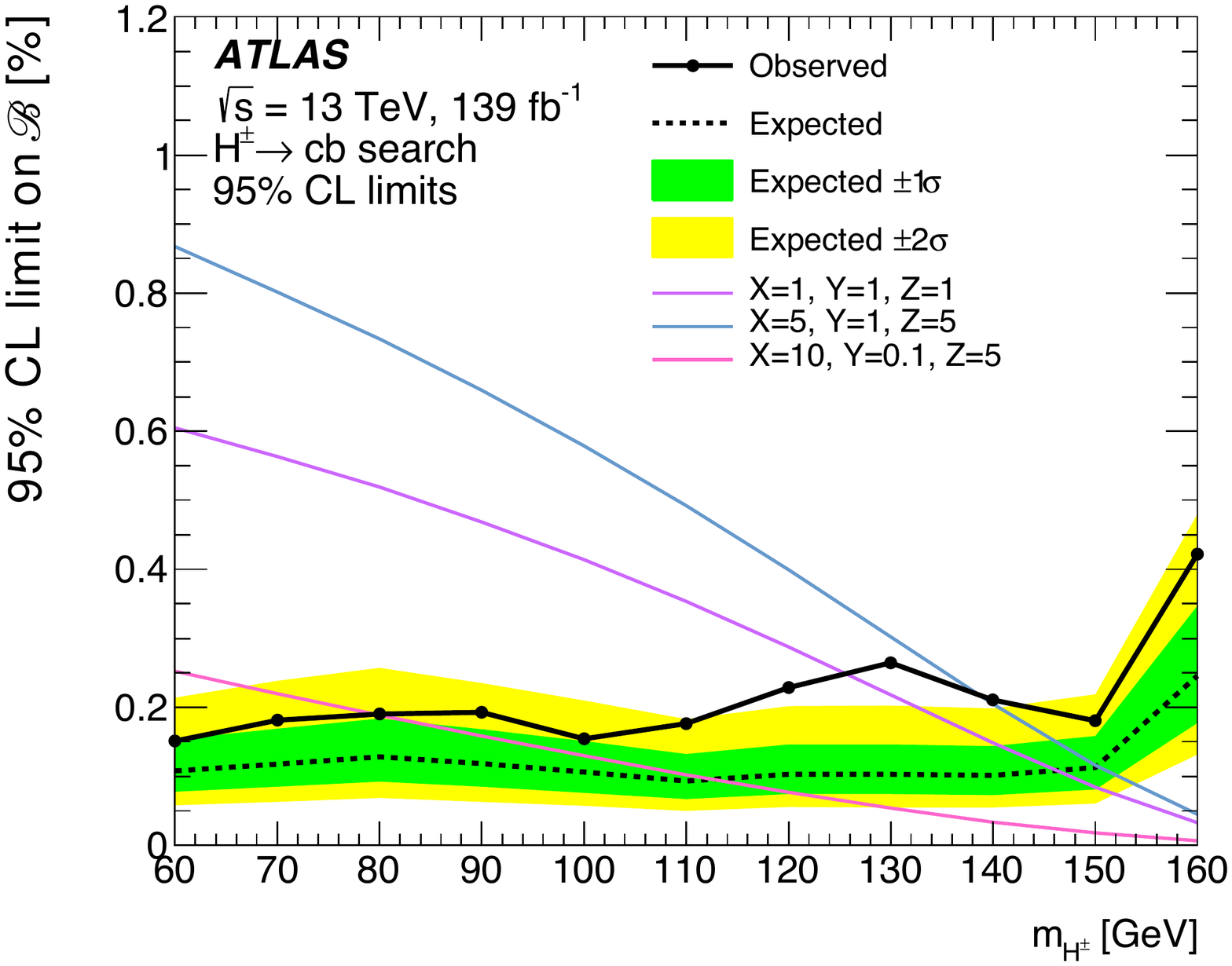}}
\caption{The observed (solid) 95\% CL upper limits on $\BR=\BR(t\rightarrow H^{\pm}b) \times \BR(H^{\pm}\rightarrow cb)$ as a function of \hpm\ and the expectation (dashed) under the background-only hypothesis. The inner green and outer yellow shaded bands show the $\pm1\sigma$ and $\pm2\sigma$ uncertainties of the expected limits.
The exclusion limits are presented for \hpm\ between 60 and 160~\gev\ with 10~\gev\ \hpm\ spacing and linear interpolation between adjacent mass points.
Superimposed on the upper limits, the predictions from the 3HDM~\cite{Akeroyd:2016ssd,Akeroyd:2018axd} are
shown, corresponding to three benchmark values for the parameters $X$, $Y$, and $Z$ described in the text.
}
\label{fig:limits}
\end{center}
\end{figure*}
 
There is no significant excess of data events above the background expectation, and 95\% CL limits are set on
the product of branching fractions $\BR$.
Figure~\ref{fig:limits} shows the observed (expected) 95\% CL upper limits on the branching fraction
$\BR$ as a function of \hpm;
they range from
0.15\% (0.09\%) up to 0.42\% (0.25\%)
depending on \hpm.
The acceptance loss for the $b$-jet produced from the decay \topdec increases for \hpm\ close to the top-quark mass, resulting in weaker exclusion limits.
Superimposed on the upper limits, the predictions from the 3HDM~\cite{Akeroyd:2016ssd,Akeroyd:2018axd} are
shown, corresponding to three benchmark values for the parameters $X$, $Y$, and $Z$, which are functions of the Higgs-doublet vacuum expectation values and the mixing angle between the charged Higgs bosons.
 
The observed exclusion limits are consistently weaker than the expectation.
The largest excess in data has a local significance of about $3\sigma$ for $\hpm=130$~\gev. The corresponding best-fit $\BR$ is measured to be $(0.16\pm0.06)$\%.
 
Pseudo-experiments from simulated events are generated to estimate the global significance of this excess.
In the first step, the correlation across all bins of the NN score distributions for all \hpm hypotheses is estimated using the MC samples.
Next, the pseudo-data are generated by fluctuating all NN score distribution bins corresponding to the pre-fit background expectation within their statistical uncertainty while
preserving the correlations across all \hpm values.
Each generated pseudo-experiment is analysed with the test statistic described in Section~\ref{sec:result} built for each of the considered \hpm hypothesis.
The global $p$-value is then calculated as the fraction of times the maximum local significance in each pseudo-experiment exceeds the maximum local significance observed in data.
This procedure leads to a global significance of $(2.46 \pm 0.05)\sigma$ in the considered \hpm\ range, with the uncertainty
originating from the pseudo-experiments' sample size.
 
Studies using pseudo-data, defined as the sum of all predicted post-fit backgrounds plus an injected signal of variable strength, were performed to validate the behaviour of the exclusion limits as a function of \hpm.
It was concluded that a broad excess in the considered \hpm\ range is consistent with the \hpm\ resolution, which is significantly worse than the expected dijet mass resolution ($\sim$$15\%$) due to the ambiguity in choosing the correct $b$-jet to pair with the fourth jet to reconstruct the \hpm\ mass.



\FloatBarrier
\section{Conclusion}

A search for the \hpdec decay mode in top-quark decays is presented. The search uses a dataset of $pp$ collisions collected at a
centre-of-mass energy $\sqrt{s}=13\;\tev$  between 2015 and 2018 with the ATLAS detector at CERN's Large Hadron Collider, amounting to an integrated luminosity of 139~fb$^{-1}$.
The analysis focuses on a data sample enriched in top-quark pair production, where one top quark decays into a leptonically decaying $W$ boson and a bottom quark, and the other top quark may decay into a $H^{\pm}$ boson and a bottom quark.
The search exploits the high multiplicity of $b$-jets, as expected from signal events, and deploys a neural network classifier that uses the kinematic differences between the signal and the background.
In the absence of a significant excess of data events above the background expectation, model-independent exclusion limits at 95\% confidence level on the
product of branching fractions $\BR=\BR(t\rightarrow H^{\pm}b)\,\times \BR(H^{\pm}\rightarrow cb)$ are reported as a function of \hpm.
The observed (expected) limits vary between
0.15\% (0.09\%) and 0.42\% (0.25\%)
for \hpm between 60 and 160~\gev.
The largest excess in data has a significance of about $3\sigma$ for $\hpm=130$~\gev.
In the considered \hpm\ range the global significance is estimated to be about 2.5$\sigma$.

Thanks to a much larger dataset and improved analysis techniques,
this search has an expected sensitivity to \hpdec\ in top-quark decays that is a factor of five higher than achieved in a previous search by the CMS Collaboration and explores an extended \hpm\ range.


\section*{Acknowledgments}


We thank CERN for the very successful operation of the LHC, as well as the
support staff from our institutions without whom ATLAS could not be
operated efficiently.
 
We acknowledge the support of
ANPCyT, Argentina;
YerPhI, Armenia;
ARC, Australia;
BMWFW and FWF, Austria;
ANAS, Azerbaijan;
CNPq and FAPESP, Brazil;
NSERC, NRC and CFI, Canada;
CERN;
ANID, Chile;
CAS, MOST and NSFC, China;
Minciencias, Colombia;
MEYS CR, Czech Republic;
DNRF and DNSRC, Denmark;
IN2P3-CNRS and CEA-DRF/IRFU, France;
SRNSFG, Georgia;
BMBF, HGF and MPG, Germany;
GSRI, Greece;
RGC and Hong Kong SAR, China;
ISF and Benoziyo Center, Israel;
INFN, Italy;
MEXT and JSPS, Japan;
CNRST, Morocco;
NWO, Netherlands;
RCN, Norway;
MEiN, Poland;
FCT, Portugal;
MNE/IFA, Romania;
MESTD, Serbia;
MSSR, Slovakia;
ARRS and MIZ\v{S}, Slovenia;
DSI/NRF, South Africa;
MICINN, Spain;
SRC and Wallenberg Foundation, Sweden;
SERI, SNSF and Cantons of Bern and Geneva, Switzerland;
MOST, Taiwan;
TENMAK, T\"urkiye;
STFC, United Kingdom;
DOE and NSF, United States of America.
In addition, individual groups and members have received support from
BCKDF, CANARIE, Compute Canada and CRC, Canada;
PRIMUS 21/SCI/017 and UNCE SCI/013, Czech Republic;
COST, ERC, ERDF, Horizon 2020 and Marie Sk{\l}odowska-Curie Actions, European Union;
Investissements d'Avenir Labex, Investissements d'Avenir Idex and ANR, France;
DFG and AvH Foundation, Germany;
Herakleitos, Thales and Aristeia programmes co-financed by EU-ESF and the Greek NSRF, Greece;
BSF-NSF and MINERVA, Israel;
Norwegian Financial Mechanism 2014-2021, Norway;
NCN and NAWA, Poland;
La Caixa Banking Foundation, CERCA Programme Generalitat de Catalunya and PROMETEO and GenT Programmes Generalitat Valenciana, Spain;
G\"{o}ran Gustafssons Stiftelse, Sweden;
The Royal Society and Leverhulme Trust, United Kingdom.
 
The crucial computing support from all WLCG partners is acknowledged gratefully, in particular from CERN, the ATLAS Tier-1 facilities at TRIUMF (Canada), NDGF (Denmark, Norway, Sweden), CC-IN2P3 (France), KIT/GridKA (Germany), INFN-CNAF (Italy), NL-T1 (Netherlands), PIC (Spain), ASGC (Taiwan), RAL (UK) and BNL (USA), the Tier-2 facilities worldwide and large non-WLCG resource providers. Major contributors of computing resources are listed in Ref.~\cite{ATL-SOFT-PUB-2023-001}.


\FloatBarrier
\printbibliography

\clearpage

\end{document}